\documentclass{nature}

\pdfoutput=1

\usepackage[margin=0.7in]{geometry}

\usepackage{graphicx}
\usepackage{epstopdf}
\usepackage{amssymb}
\usepackage{mathtools}
\usepackage{xfrac}
\usepackage{color}
\usepackage{lineno}

\usepackage{siunitx}
\usepackage{tabularx}
\usepackage{longtable}
\usepackage{comment}
\usepackage{romannum}
\usepackage{ulem}
\usepackage{caption}
\usepackage{subcaption}

\def \d{\mathrm{d}}
\def \ls{\ell_{\textrm{sat}} }
\def \lss{\ell_{\textrm{sat}}^{\textrm{sub}} }

\definecolor{red}{rgb}{1,0,0}

\title{Coevolving aerodynamic and impact ripples on Earth
 }

\author{Hezi Yizhaq$^{1^\ast}$, Katharina Tholen$^{2^\ast}$, Lior Saban$^{3}$, Nitzan Swet$^{3}$, Conner Lester$^4$, Simone Silvestro$^{5,6}$, Keld R.  Rasmussen$^7$,  Jonathan P. Merrison$^8$,  Jens J. Iversen$^8$,  Gabriele Franzese$^5$,  Klaus Kroy$^{2}$,   Thomas P\"ahtz$^{9,10^\dag}$, Orencio Dur\'an$^{11^\dag}$, Itzhak Katra$^{3^\dag}$}

\begin{document}

\maketitle
\begin{affiliations}
 \item Department of Solar Energy and Environmental Physics, Ben-Gurion University of the Negev, Be'er Sheva, Israel
  \item Institute for Theoretical Physics, Leipzig University, Brüderstraße 16, 04103 Leipzig, Germany
   \item Department of Environmental, Geoinformatics, and Urban Planning Sciences, Ben-Gurion University of the Negev, Be'er Sheva, Israel
  \item Division of Earth and Ocean Sciences, Duke University, Box 90227, Durham, 27708-0227, North Carolina, USA
  \item Istituto Nazionale di Astrofisica, Osservatorio di Capodimonte, Napoli, Italy.
	\item Carl Sagan Center, SETI Institute, Mountain View, CA, USA.
	 \item Department of Earth Sciences, Aarhus University, 8000 Aarhus C, Denmark
  \item Institute for Physics and Astronomy, Aarhus University, 8000 Aarhus C, Denmark
 \item Institute of Port, Coastal and Offshore Engineering, Ocean College, Zhejiang University, 316021 Zhoushan, China
 \item Donghai Laboratory, 316021 Zhoushan, China
   \item Department of Ocean Engineering, Texas A\&M University, College Station, Texas 77843-3136, USA
\end{affiliations}
$^\ast$ These authors contributed equally to this work
\newpage

\pagenumbering{arabic}

\begin{abstract}
Windblown sand creates multiscale bedforms on Earth, Mars, and other planetary bodies. According to conventional wisdom, decameter-scale dunes and decimeter-scale ripples emerge via distinct mechanisms on Earth: a hydrodynamic instability related to a phase shift between the turbulent flow and the topography, and a granular instability related to a synchronization of hopping grains with the topography. Here, we report the reproducible creation of coevolving centimeter and decimeter-scale ripples on fine-grained monodisperse sand beds in ambient-air and low-pressure wind-tunnels, revealing two adjacent mesoscale growth instabilities. Their morphological traits and our quantitative grain-scale numerical simulations authenticate the smaller structures as impact ripples but point at a hydrodynamic origin for the larger ones. This suggests that the aeolian transport layer would have to partially respond to the topography on a scale comparable to the average hop length, hence faster than previously thought, but consistent with the phase lag of the inferred aeolian sand flux relative to the wind. Hydrodynamic modelling supports the existence of hydrodynamic aerodynamic ripples on Earth, connecting them mechanistically to megaripples and to the debated Martian ripples. We thereby propose a unified framework for mesoscale granular bedforms found across the Solar System.

\end{abstract}


\begin{figure}[]
\centering
\includegraphics{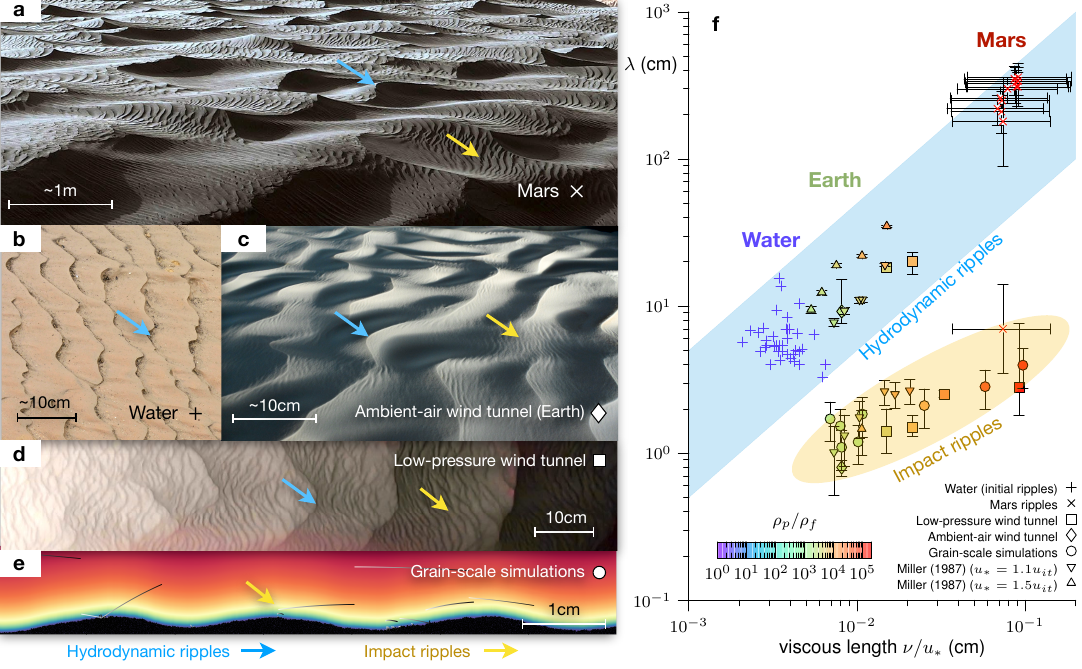}
\caption{
\footnotesize
\textbf{Examples and classification of mesoscale bedforms for a wide range of environmental conditions.}
Meter-scale ripples (of disputed origin\cite{Ewingetal17,Bakeretal18b,Siminovichetal19,Sullivanetal20, lorenz2020martian,Lapotreetal16,Lapotreetal18,Duranetal19, Lapotreetal21, yizhaq2021turbulent,rubanenko2022distinct,silvestro2016dune}) on Mars (\textbf{a}) morphologically resemble terrestrial subaqueous ripples (\textbf{b}),  but are decorated with secondary, centimeter-scale ripples. Contrary to conventional wisdom, our ambient-air (\textbf{c}) and low-pressure (\textbf{d}) wind-tunnel experiments produce coevolving superimposed centimeter- and decimeter-scale ripples,  on Earth, in agreement with previous data from the Martian surface wind tunnel MARSWIT\cite{miller1987wind}. Together with our grain-scale numerical simulations (\textbf{e}) and previous measurements (orbital and in-situ data for Mars\cite{Lapotreetal16}, selected initial subaqueous ripples\cite{Duranetal19} and the MARSWIT data\cite{miller1987wind}), our wind tunnel data fall into two distinct categories  (\textbf{f}) when plotting their wavelengths $\lambda$ as a function of the viscous sublayer scale $\nu/u_\ast$, with $\nu$ the kinematic viscosity and $u_\ast$ the wind friction velocity.  The blue-shaded area indicates the typical range of initial wavelengths of  hydrodynamic (subaqueous, aerodynamic, low-pressure) ripples suggested by scaling arguments\cite{Lapotreetal16,lapotre2017sets,Lapotreetal21}, but excluded by previous mechanistic morphodynamic modeling\cite{Duranetal19} for Earth (see text). The orange-shaded area guides the eye to a group of smaller bedforms consistent with impact-ripple numerical simulations ($\circ$). The color of the symbols represents the particle-fluid density ratio $\rho_p/\rho_f$.  See Methods for more details and Supplementary Tables S1-S3 for data. 
}
\label{introduction}
\end{figure}

Sand waves spontaneously created by atmospheric flows are widely observed on diverse celestial bodies\cite{hayes2018dunes,Bourkeetal10,Rasmussenetal15,Diniegaetal17}. Their shapes and sizes, and even their classification, vary as a function of atmospheric density, pressure, and turbulent flow properties\cite{Charruetal13}  (Fig.~1), making it difficult to infer the physical origins of extraterrestrial bedforms via analogies to their intensely studied counterparts on Earth\cite{Telferetal18}. In particular, there is an ongoing debate about two mesoscale Martian bedforms in well-sorted  fine  sands  (so-called small and large ripples, see Fig.~1a)\cite{Ewingetal17,Bakeretal18b,Siminovichetal19,Sullivanetal20, lorenz2020martian,Lapotreetal16,Lapotreetal18,Duranetal19, Lapotreetal21, yizhaq2021turbulent,rubanenko2022distinct,silvestro2016dune}. While it is generally agreed upon that small Martian ripples are impact ripples, there are currently competing hypotheses regarding the physical origin of their larger siblings, and also of so-called transverse aeolian ridges (TARS)\cite{Bridgesetal12b,balme2008transverse,zimbelman2010transverse,hugenholtz2017morphology} or megaripples\cite{Sullivanetal08,  jerolmack2006spatial,weitz2018sand}. Are they mature impact ripples that evolve through growth and mergers of smaller ones\cite{Siminovichetal19,Sullivanetal20,lorenz2020martian,yizhaq2021turbulent}? Or does their strong morphological resemblance to subaqueous terrestrial ripples (Fig.~1b) attest to their hydrodynamic origin\cite{Lapotreetal16,Lapotreetal18,Duranetal19,Lapotreetal21}? The latter hypothesis can indeed be supported by a recent model that unifies hydrodynamic bedform evolution across subaqueous and planetary atmospheric environments\cite{Duranetal19}.  It arguably allows for two coexisting types of hydrodynamic instabilities, consistent with dunes and hydrodynamic ripples  (also known as ``fluid-drag ripples"\cite{Bagnold41,southard1990bed,greeley1987wind, Lapotreetal18,Lapotreetal21}, ``aerodynamic ripples"\cite{Wilson72} or ``wind-drag ripples''\cite{Lapotreetal16}), respectively, separated due to a hydrodynamic anomaly\cite{Charruetal13, abrams1985relaxation,frederick1988velocity,claudin2017dissolution}  suppressing the growth of bedforms with intermediate wavelengths (cf. Fig.~2b in Ref.\cite{Duranetal19}). However, this requires sufficiently viscous conditions at the grain scale, as naturally realized for sand transport by subaqueous flows or the highly viscous low-pressure winds on Mars,  suggesting that ordinary terrestrial laboratory studies are poorly suited to resolve the debate\cite{Duranetal19,Lapotreetal21,Sullivanetal20}.  Indeed, most studies of uniform sands have reported only one terrestrial bedform smaller than dunes, identified as impact ripple\cite{ellwood1975small,Walker81,Schmerleretal16,Chengetal18, Andreottietal06,Sharp63,Rasmussenetal15}. Yet this appears to be somewhat at odds with occasional observations of sinusoidal crest lines and other morphological similarities with subaqueous ripples\cite{cornish1897formation,Bagnold41,Wilson72, Lapotreetal18,greeley1987wind}. Could these, together with casual mentions of ``secondary ripples"\cite{SeppalaLinde78,Chengetal18} or  a ``sudden morphological shift'' of ripples\cite{Walker81,greeley1987wind,Bagnold41,ling1998wind,Chengetal18}, in fact hint at the co-existence of two distinct terrestrial mesoscale bedforms, as some authors have speculated before \cite{Bagnold41,Wilson72,greeley1987wind,Lapotreetal16, Lapotreetal18}? This is indeed supported by Figs.~1c and d,  and by a little known historical NASA Memo from 1987 about the Martian Surface Wind Tunnel (MARSWIT) experiments\cite{miller1987wind}.

\section{Two coevolving ripple populations}
\begin{figure}[]
\centering
\includegraphics{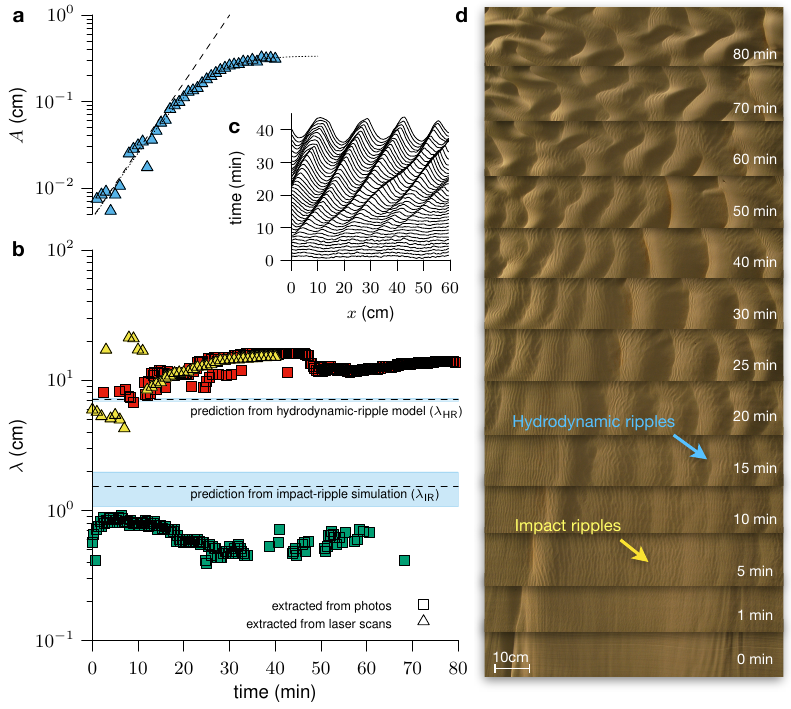}
\caption{
\footnotesize
\textbf{Multiple growth instabilities in ambient-air wind-tunnel measurements} (grain size $d\approx 90~\mathrm{\mu m}$ and wind friction velocity $u_\ast\approx0.184~ \mathrm{ m/s}$). Amplitude $A$ (\textbf{a}, blue) and wavelength $\lambda$ (\textbf{b}, red and yellow) of decimeter-scale ripples (\textbf{c}), and wavelength of the superimposed centimeter-scale ripples (\textbf{b}, green), emerging simultaneously from an initially flat sand bed under a constant wind  (\textbf{d}).  The inset (\textbf{c}) is a spatiotemporal plot showing the emergence of decimeter-scale ripples along a transect in the central section of the wind tunnel ($\lambda$ shown in \textbf{b}, yellow), while the time-lapse photos in (\textbf{d}) show the full evolution of both ripple scales ($\lambda$ shown in \textbf{b}, red and green). The notion of two distinct linear growth instabilities is supported by the initial exponential growth of the larger ripples' amplitudes (dashed line in \textbf{a}), as inferred from an empirical fit of the whole time series (dotted line, see Methods),  at well-separated, invariable initial ripple wavelengths (\textbf{b}), in accord with the predicted initial wavelengths (dashed lines) from our aeolian impact-ripple numerical simulations (Fig.~1f, Methods and Supplementary Information) and from the hydrodynamic-ripple model (Fig.~4).  Extracted wavelengths (amplitudes) are noisy,  for the large ripples at early times ($\lesssim 10 ~\mathrm{min}$) because of the superimposed small ripples (Methods), and for the small ripples at late times ($\gtrsim 40~\mathrm{min}$) because of disturbances due to the emerging larger ripples, which may even cause their partial flattening (\textbf{d}).  The wrinkles at $t=0~\mathrm{min}$ in the time-lapse photos (\textbf{d}) are minor (inconsequential) imperfections in the initial bed preparation.  
}
\label{linear_instability}
\end{figure}

To substantiate this central hypothesis, we carried out systematic ambient-air and low-pressure wind-tunnel experiments, as described in Methods.  Flat sand beds were prepared at the bottom of the tunnel's test section and subjected to diverse steady wind regimes. Thereby we could identify conditions near the aeolian transport threshold that reproducibly allow for the simultaneous evolution of two superimposed centimeter- and decimeter-scale bedforms (Figs.~1c, d and Movie S1). Their initially exponentially growing amplitudes (Fig.~2a) at almost constant wavelengths (Fig.~2b) strongly suggest two distinct linear growth instabilities rather than a nonlinear merging phenomenon or wind-strength dependent metamorphosis\cite{greeley1987wind,Bagnold41}. This finding thus strongly  substantiates and rationalizes the above-mentioned circumstantial evidence.  Similar to centimeter-scale structures reported in previous studies\cite{Bagnold41,SeppalaLinde78, Wilson72,Lapotreetal18,miller1987wind}, the rapidly growing smaller bedforms have a wrinkled,  stripe-like appearance, and their spacing and orientation seem to follow the local granular transport geometry, modulated by the larger bedforms (Figs.~1c, d and 2d). Local periodicity, as, e.g.,  seen on the lee side of the larger bedforms,  seems to be a key trait, in line with the picture of impact ripples as a phase-locking resonance between bed corrugations and the hopping grains\cite{Bagnold41,Duranetal14b}.  In contrast, the more slowly emerging wavetrains of larger bedforms are visually quite distinct, and the lateral sinuosity of their smooth crest lines seems to follow the geometry of the wind field rather than the local grain hopping (Fig.~1c, d and 2d). This supports the notion of a hydrodynamic instability far below the dune scale, creating, under suitable conditions, terrestrial aeolian ripples analogous to subaqueous ripples\cite{Fourriereetal10,ColombiniStocchino11,rubin1980single,southard1990bed} and Martian large ripples\cite{silvestro2016dune,Lapotreetal16,Lapotreetal18} (Fig.~1).

From a more quantitative point of view, we find a close correspondence between the smaller (suspected impact) ripples from (i) our ambient-air and low-pressure wind tunnel and (ii) the MARSWIT experiment\cite{miller1987wind}, and (iii) small ripples on Mars\cite{Lapotreetal16}. The similarity extends to the sole bedform produced in our grain-scale numerical simulations of aeolian saltation (Figs.~1f and S4, Methods and Supplementary Information, Movies S2 and S3), and this comprises their fast evolution, appearance, and typical wavelengths (roughly on the order of the grains' hop length). By construction, our simulations cannot produce hydrodynamic bedforms, since they neglect topographic feedback on the driving flow. If expressed in natural units (i.e., the viscous sublayer scale $\nu/u_\ast$, with $\nu$ the kinematic fluid viscosity and $u_\ast$ the wind friction velocity), the rescaled wavelengths $\lambda_{\rm IR} u_\ast/\nu= \mathcal{O} (10^2)$ of all suspected impact ripples indeed fall into a narrow corridor (Fig.~1f, orange shaded area), clearly separated from the domain of the suspected hydrodynamic bedforms (Fig.~1f, blue shaded area). The scale separation extends to the growth rates: the smaller ripples in our experiments matured before the large ones even became discernible. In summary, since two bedforms of unequal size and scaling cannot emerge via the same (linear) growth instability,  our above findings strongly support our central hypothesis. Strikingly, the rescaled wavelengths $\lambda_{\rm HR} u_\ast/\nu = \mathcal{O} (10^3)$ of the larger ripples in our wind tunnel experiments and in the MARSWIT study\cite{miller1987wind} exceed the average hop length quite significantly and fall into the theoretically  plausible corridor for hydrodynamic ripples in the hydrodynamically smooth regime\cite{Duranetal19,Lapotreetal16,lapotre2017sets,Lapotreetal21}  with aerodynamic roughness $z_0\lesssim\nu/u_\ast$ (Methods), spanned by subaqueous ripples\cite{grazer1982experimental,yalin1985determination} and large ripples on Mars\cite{Lapotreetal16}, respectively (Fig.~1f, blue shaded area). 

\section{Subscale saturation}
\begin{figure}[]
\centering
\includegraphics{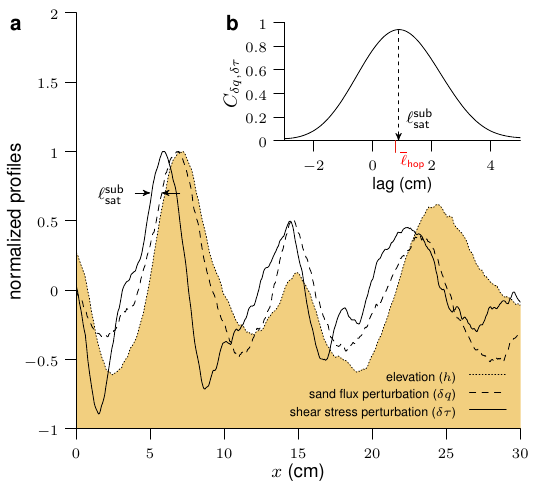}
\caption{
\footnotesize
\textbf{Estimating the subscale saturation length $\ell_{\rm sat}^{\rm sub}$ under ambient-air conditions.}
(\textbf{a}) Exemplary bed elevation profile $h(x)$,  experimentally inferred sand flux perturbation $\delta q(x)$ (Methods), and  calculated bed shear stress pertubation $\delta \tau (x)$ (Methods),  all normalized by their maximum values,  during the initial formation of large ripples in the center of the ambient-air wind tunnel ($t = 11~{\rm min}$, see Fig.~2). The sand flux is shifted relative to the bed shear stress by a well-defined lag distance, quantified by the maximum of the cross-correlation function  $C_{\delta q,\delta \tau}$ (\textbf{b}, see Methods). This lag is about $50$ times smaller than previous estimates obtained on much coarser scales\cite{Elbelrhitietal05,Andreottietal10,Selmanietal18,lu2021direct}, and thus indicative of a partial, subscale saturation over a length $\ell_{\rm sat}^{\rm sub}$ on the order of the average hop length $\bar{\ell}_{\rm hop} \approx 0.8~{\rm cm}$ (calculated from grain-scale simulations, see Supplementary Information). }
\label{l_sat}
\end{figure}
Taken together, the above observations strongly support the notion of a hydrodynamic instability far below the minimum dune size, thereby restoring the plausible\cite{Bagnold41,Wilson72,greeley1987wind,Lapotreetal16, Lapotreetal18} but previously disputed\cite{Duranetal19,Lapotreetal21,Sullivanetal20} symmetry between bedform evolution on Mars, Earth, and under water. 
This may seem surprising from a mechanistic perspective\cite{Duranetal19}, since the existence of hydrodynamic ripples requires a sufficiently small saturation length $\ell_{\rm sat}$ (the characteristic response length of aeolian sand transport to wind perturbations\cite{Sauermannetal01,Duranetal11}), roughly $\ell_{\rm sat}u_\ast/\nu\lesssim10^3$. Such conditions may indeed seem difficult to satisfy for the low-viscosity atmosphere on Earth ($\nu\approx1.5\times10^{-5}~\mathrm{m^2/s}$), even for our fine sand and low wind speed, since decimeter- and meter-scale measurements suggest 
$\ell_{\rm sat}\approx4500d$\cite{Elbelrhitietal05,Andreottietal10,Selmanietal18, lu2021direct} ($\ell_{\rm sat}u_\ast/\nu\approx5000$ for $d\approx90~\mu\mathrm{m}$ at $u_\ast \approx 0.184~\mathrm{ m/s}$), in apparent conflict with the above cumulative evidence.\\
As a way to resolve this conundrum, we postulate the existence of a secondary, subscale modulation of the aeolian transport layer.  
This rather unconventional hypothesis can directly be tested by comparing the modulation of the sand flux $\delta q(x)$,  as extracted from our ambient-air wind-tunnel measurements (Methods), with the calculated fluid shear stress modulation $\delta \tau(x)$ (Fig.~3a).  As expected for a hydrodynamic growth instability,  we find a strong correlation, with a lag-distance that is indeed indicative of a subscale saturation length $\ell_{\rm sat}^{\rm sub}\approx \bar{\ell}_{\rm hop}$, substantially below the saturation length $\ell_{\rm sat}\approx 50 \bar{\ell}_{\rm hop}$ of the total flux (Fig.~3b).  Physically,  a partial subscale saturation hints at multiple relaxation processes, presumably associated with a  stratification within the aeolian sand transport layer, as indeed traditionally invoked as a key ingredient in impact-ripple models\cite{Duranetal14b,Bagnold41,Anderson87}, analytical saltation models\cite{Andreotti04,Lammeletal12,Lammeletal17, tholen2023anomalous}, and empirically supported by studies of megaripples\cite{Bagnold41,AndersonBunas93,Lammeletal18, tholen2022megaripple, Yizhaq04,Makse00,ManukyanPrigozhin09} (oversized ripples made of bimodal sands). 

\begin{figure}[]
\centering
\includegraphics{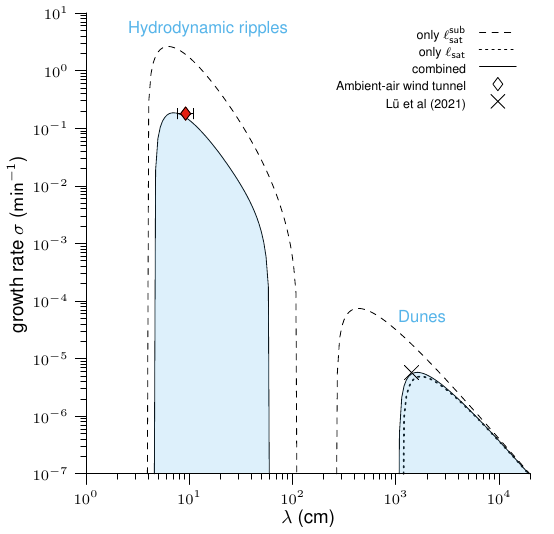}
\caption{
\footnotesize
\textbf{Model predictions for ambient-air conditions.} Integrating an additional subscale relaxation process into a state-of-the-art morphodynamic model\cite{Duranetal19} (Methods) leads to a new dispersion relation $\sigma(\lambda)$ (solid line).  For terrestrial conditions, it predicts a hydrodynamic ripple mode in addition to the classical dune mode associated with the overall saturation process, in good agreement with our experimental data for incipient large-ripples  ($\diamond$, $t\lesssim 15 ~{\rm min}$, see Fig.~2 and Methods) and field measurements of incipient sand dunes\cite{lu2021direct},  corrected for the different grain sizes and wind speeds (Methods). 
The dotted and dashed lines indicate two limiting scenarios considering merely one sand-flux relaxation process: conventional large-scale  saturation ($\ell_{\rm sat}\sim 0.4~{\rm m}$) {\it or} the newly postulated subscale saturation ($\ell_{\rm sat}^{\rm sub}  \sim 0.8~{\rm cm}$), alone. 
Their linear combination (solid line) involves, as the only free parameter, the fraction $\gamma \approx0.1$ of the total sand flux that is modulated at the smaller scales. It is adjusted to fit the measured growth rate of the large ripples (Fig.~2a, Methods). 
}
\label{sigma}
\end{figure}
Including this additional, subscale saturation in the state-of-the-art theoretical model of hydrodynamic bedform evolution\cite{Duranetal19}  (Methods) would make it compatible with a mesoscale hydrodynamic bedform (an aerodynamic ripple) on Earth. The ensuing integral dispersion relation would moreover  be in good quantitative agreement with the large ripples'  observed initial wavelength and growth rate 
(Fig. ~4), while preserving the empirically well established growth instability for dunes\cite{Charruetal13,Duranetal19,Fourriereetal10,lu2021direct}. This further supports the hydrodynamic-origin hypothesis and again hints at a second, subscale saturation process.
Additionally,  the new interpretation implicates an important role of the bounded stability range, adjacent to the hydrodynamic anomaly\cite{Charruetal13,abrams1985relaxation,frederick1988velocity, claudin2017dissolution}
which stabilizes the ripple amplitude and wavelength\cite{Duranetal19,rubanenko2022distinct}.  Although the quantitative details remain to be worked out, such amended hydrodynamic model could then naturally account for the observed initial wavelength selection for large fine-grained terrestrial ripples\cite{Andreottietal06}, their final stabilization\cite{Bagnold41,Walker81,SeppalaLinde78,Andreottietal06,Chengetal18,Rasmussenetal15}, and possibly (as in the subaqueous case\cite{Duranetal19}) their sudden flattening under strong flows\cite{Walker81,ling1998wind,Sharp63,Schmerleretal16,Bagnold41}.

\section{Discussion}

Generally speaking, the study of the coevolution of the implicated two types of ripples is complicated by their unequal growth rates. On Earth, in contrast to Mars\cite{Duranetal19}, also their rather weak size separation (Fig.~1) hampers their distinction and makes unfavorable interactions of the two bedforms more likely\cite{Lapotreetal16,Lapotreetal18}, especially for natural sands. For example, near the transport threshold, the larger,  aerodynamic ripples take much longer to evolve and mature than the smaller, impact ripples but may eventually disturb their periodicity (Fig.~2). 
Therefore, depending on the explicit setup (e.g., run time, wind speed, saturation, wind-tunnel geometry, or sand characteristics), one or the other ripple type may easily elude detection, which could be a reason why only few terrestrial studies have clearly discriminated the two types of ripples in the past\cite{Bagnold41,SeppalaLinde78,miller1987wind, Wilson72,Lapotreetal18,greeley1987wind}.  While it is tempting to speculate that Bagnold could have discovered the coevolution of both ripple types, if he had run his measurements at low wind speed longer, corresponding systematic field observations, beyond suggestive accidental observations (see, e.g., Fig. 2d-e in Ref.\cite{Lapotreetal18}), ought to be quite challenging,  in particular in view of the predominant dune-sand composition ($d\approx250 \, \mathrm{\mu m}$\cite{pye2008aeolian, lancaster2013geomorphology},  probably outside the hydrodynamically smooth regime\cite{Duranetal19}). It lets one expect an even richer phenomenology with multiple interacting mechanisms, rendering model predictions potentially contingent on minor grain-scale details (grain size, grain shape) as well as on the wind speed and saturation history, moisture, etc.
Finally,  the possible scenario of natural wind-strength fluctuations promoting multiple wavelengths simultaneously\cite{Sharp63,Wilson72} could also encumber the identification of terrestrial hydrodynamic ripples merely by their size and occasional bedform superpositions.

Coming finally back to more polydisperse sands,  broad grain size distributions can substantially blur the ideal picture sketched so far.  Intriguingly, they allow for yet another decimeter- to meter-scale bedform\cite{Bagnold41,Yizhaqetal09,Sharp63,ellwood1975small,Lammeletal18}, namely so-called megaripples. Their unique bimodal surface grain-size distribution allows them to grow larger than the (size-limited) hydrodynamic ripples made of unimodal sands, from which they  might gradually emerge by progressive sand sorting\cite{KatraYizhaq17,Bagnold41,Sharp63, McKennaNeumanBedard17}. Without closer inspection, both ripple types may easily be mistaken for mature impact ripples\cite{AndersonBunas93, Yizhaq04,Makse00,ManukyanPrigozhin09}. While  in line with one of the hypotheses for large-ripple formation on Mars\cite{Siminovichetal19,Sullivanetal20,lorenz2020martian}, such interpretation could hardly account for the barchanoid shapes and other morphological traits of megaripples, indicative for a hydrodynamic growth mechanism\cite{Lammeletal18}, plus their lead position in the known size hierachy of mesoscale bedforms\cite{Sharp63,Wilson72,Lammeletal18},  which is particularly pronounced on Mars \cite{Sullivanetal08,Lapotreetal16,Lapotreetal18}.
Hence,  their abundance, with often conspicuously dune-like traits\cite{Yizhaqetal09,Sharp63,yizhaq2012a}, may currently be the least disputable field evidence for the subscale hydrodynamic growth instability invoked to explain our present observations. 

\section{Conclusions and Outlook}
In conclusion, we have reported ambient-air and low-pressure wind tunnel experiments showing two coevolving mesoscale bedforms.  We find the often overlooked centimeter-scale ripples to be consistent with our grain-scale simulations of impact ripples and the decimeter-scale ripples  with a hydrodynamic growth instability akin to that established for water\cite{Fourriereetal10,Charruetal13} and recently proposed for Mars\cite{Duranetal19,Lapotreetal16,lapotre2017sets,Lapotreetal21}. This calls for a revision of the conventional classification of mesoscale aeolian bedforms in favor of a more coherent framework that includes the existence of aerodynamic ripples (i.e., of hydrodynamic origin) on Earth, prompting a reappraisal and synthesis of  previous  explanations for impact ripples\cite{Duranetal14b}, megaripples\cite{Lammeletal18}, subaqueous bedforms\cite{Duranetal19,Fourriereetal10,Charruetal13, Lapotreetal16,lapotre2017sets,Lapotreetal21}, as well as small and large Martian ripples\cite{Lapotreetal16,lapotre2017sets,Lapotreetal18,Duranetal19,Lapotreetal21}.  It would restore a satisfying symmetry between bedform evolution on Mars, Earth, and in water\cite{Lapotreetal16,lapotre2017sets,Lapotreetal21}, and naturally allow for a smooth crossover between large ripples and megaripples.  

As the core element of the anticipated theoretical framework, we have pinpointed a hitherto unexplored mesostructure of the aeolian transport layer that admits an additional, partial saturation process, quantified by a small (subscale) saturation length $\ell_{\rm sat}^{\rm sub} \ll \ell_{\rm sat}$ on the order of the average hop length. 
While a predictive quantitative model for it remains presently elusive, our findings stipulate ample future investigations.  For example,  can the postulated subscale saturation process be evidenced directly? If so, can a corresponding stratification of the aeolian transport layer be demonstrated to predict conditions for the coexistence of two saturation processes?

Our proposed paradigm change implies that earlier literature reports may easily have confounded impact ripples with aerodynamic  ripples and reported hybrid characteristics.  The associated lack of systematic dedicated measurements of both ripples types in turn raises a number  of further questions: When and why do actual impact or aerodynamic  ripples flatten? How do  they evolve as a function of environmental conditions (e.g., wind speed and grain size)? How strongly are they affected by mutual interference,  nonlinearities, or grain size, grain shape, and wind variability? And,  how can they unambiguously be discerned under natural conditions, in the field?
Our findings clearly call for a wide concerted experimental and theoretical research effort, into which we hope to entice a broad, interdisciplinary research community.


\begin{addendum}

 \item This research was supported by the Israel Science Foundation ISF (no. 1270/20) for I.K., and by the German-Israeli Foundation for Scientific Research and Development  GIF (no. 155-301.10/2018) for I.K. and K.K..  
 T.P. acknowledges support from grant Young Scientific Innovation Research Project of Zhejiang University (no. 529001*17221012108). O.D.V. acknowledges support from the Texas A\&M Engineering Experiment Station.  This work has been funded by Europlanet grant no. 871149 (Project number: 20-EPN-054) for S.S., K.R.R., J.P.M., and G.F. Europlanet 2024 RI has received funding from the European Union’s Horizon 2020 research and innovation program.
 
 \item[Author contributions] I.K., H.Y., L.S., and N.S. designed and conducted the ambient-air wind-tunnel experiments; K.T. devised the theoretical approach; O.D. and K.T. performed the theoretical hydrodynamic bedform modeling; C.L. conducted the impact-ripple simulations; T.P.  conducted the grain-scale transport simulations; H.Y., I.K., S.S., K.R.R., J.P.M., J.J.I. and G.F. designed and conducted the low-pressure wind-tunnel experiments; O.D. and K.T. analyzed the data; K.T., K.K., T.P, O.D.  wrote the paper. All authors discussed the results and implications and commented on the manuscript at all stages.

 \item[Competing Interests] The authors declare no competing interests.

\end{addendum}


\section*{Figure Captions}

\textbf{Figure 1}:\\
\textbf{Examples and classification of mesoscale bedforms for a wide range of environmental conditions.}
Meter-scale ripples (of disputed origin\cite{Ewingetal17,Bakeretal18b,Siminovichetal19,Sullivanetal20, lorenz2020martian,Lapotreetal16,Lapotreetal18,Duranetal19, Lapotreetal21, yizhaq2021turbulent,rubanenko2022distinct,silvestro2016dune}) on Mars (\textbf{a}) morphologically resemble terrestrial subaqueous ripples (\textbf{b}),  but are decorated with secondary, centimeter-scale ripples. Contrary to conventional wisdom, our ambient-air (\textbf{c}) and low-pressure (\textbf{d}) wind-tunnel experiments produce coevolving superimposed centimeter- and decimeter-scale ripples,  on Earth, in agreement with previous data from the Martian surface wind tunnel MARSWIT\cite{miller1987wind}. Together with our grain-scale numerical simulations (\textbf{e}) and previous measurements (orbital and in-situ data for Mars\cite{Lapotreetal16}, selected initial subaqueous ripples\cite{Duranetal19} and the MARSWIT data\cite{miller1987wind}), our wind tunnel data fall into two distinct categories  (\textbf{f}) when plotting their wavelengths $\lambda$ as a function of the viscous sublayer scale $\nu/u_\ast$, with $\nu$ the kinematic viscosity and $u_\ast$ the wind friction velocity.  The blue-shaded area indicates the typical range of initial wavelengths of  hydrodynamic (subaqueous, aerodynamic, low-pressure) ripples suggested by scaling arguments\cite{Lapotreetal16,lapotre2017sets,Lapotreetal21}, but excluded by previous mechanistic morphodynamic modeling\cite{Duranetal19} for Earth (see text). The orange-shaded area guides the eye to a group of smaller bedforms consistent with impact-ripple numerical simulations ($\circ$). The color of the symbols represents the particle-fluid density ratio $\rho_p/\rho_f$.  See Methods for more details and Supplementary Tables S1-S3 for data. 

\textbf{Figure 2}:\\
\textbf{Multiple growth instabilities in ambient-air wind-tunnel measurements} (grain size $d\approx 90~\mathrm{\mu m}$ and wind friction velocity $u_\ast\approx0.184~ \mathrm{ m/s}$). Amplitude $A$ (\textbf{a}, blue) and wavelength $\lambda$ (\textbf{b}, red and yellow) of decimeter-scale ripples (\textbf{c}), and wavelength of the superimposed centimeter-scale ripples (\textbf{b}, green), emerging simultaneously from an initially flat sand bed under a constant wind  (\textbf{d}).  The inset (\textbf{c}) is a spatiotemporal plot showing the emergence of decimeter-scale ripples along a transect in the central section of the wind tunnel ($\lambda$ shown in \textbf{b}, yellow), while the time-lapse photos in (\textbf{d}) show the full evolution of both ripple scales ($\lambda$ shown in \textbf{b}, red and green). The notion of two distinct linear growth instabilities is supported by the initial exponential growth of the larger ripples' amplitudes (dashed line in \textbf{a}), as inferred from an empirical fit of the whole time series (dotted line, see Methods),  at well-separated, invariable initial ripple wavelengths (\textbf{b}), in accord with the predicted initial wavelengths (dashed lines) from our aeolian impact-ripple numerical simulations (Fig.~1f, Methods and Supplementary Information) and from the hydrodynamic-ripple model (Fig.~4).  Extracted wavelengths (amplitudes) are noisy,  for the large ripples at early times ($\lesssim 10 ~\mathrm{min}$) because of the superimposed small ripples (Methods), and for the small ripples at late times ($\gtrsim 40~\mathrm{min}$) because of disturbances due to the emerging larger ripples, which may even cause their partial flattening (\textbf{d}).  The wrinkles at $t=0~\mathrm{min}$ in the time-lapse photos (\textbf{d}) are minor (inconsequential) imperfections in the initial bed preparation.  

\textbf{Figure 3}:\\
\textbf{Estimating the subscale saturation length $\ell_{\rm sat}^{\rm sub}$ under ambient-air conditions.}
(\textbf{a}) Exemplary bed elevation profile $h(x)$,  experimentally inferred sand flux perturbation $\delta q(x)$ (Methods), and  calculated bed shear stress pertubation $\delta \tau (x)$ (Methods),  all normalized by their maximum values,  during the initial formation of large ripples in the center of the ambient-air wind tunnel ($t = 11~{\rm min}$, see Fig.~2). The sand flux is shifted relative to the bed shear stress by a well-defined lag distance, quantified by the maximum of the cross-correlation function  $C_{\delta q,\delta \tau}$ (\textbf{b}, see Methods). This lag is about $50$ times smaller than previous estimates obtained on much coarser scales\cite{Elbelrhitietal05,Andreottietal10,Selmanietal18,lu2021direct}, and thus indicative of a partial, subscale saturation over a length $\ell_{\rm sat}^{\rm sub}$ on the order of the average hop length $\bar{\ell}_{\rm hop} \approx 0.8~{\rm cm}$ (calculated from grain-scale simulations, see Supplementary Information). 

\textbf{Figure 4}:\\
\textbf{Model predictions for ambient-air conditions.} Integrating an additional subscale relaxation process into a state-of-the-art morphodynamic model\cite{Duranetal19} (Methods) leads to a new dispersion relation $\sigma(\lambda)$ (solid line).  For terrestrial conditions, it predicts a hydrodynamic ripple mode in addition to the classical dune mode associated with the overall saturation process, in good agreement with our experimental data for incipient large-ripples  ($\diamond$, $t\lesssim 15 ~{\rm min}$, see Fig.~2 and Methods) and field measurements of incipient sand dunes\cite{lu2021direct},  corrected for the different grain sizes and wind speeds (Methods). 
The dotted and dashed lines indicate two limiting scenarios considering merely one sand-flux relaxation process: conventional large-scale  saturation ($\ell_{\rm sat}\sim 0.4~{\rm m}$) {\it or} the newly postulated subscale saturation ($\ell_{\rm sat}^{\rm sub}  \sim 0.8~{\rm cm}$), alone. 
Their linear combination (solid line) involves, as the only free parameter, the fraction $\gamma \approx0.1$ of the total sand flux that is modulated at the smaller scales. It is adjusted to fit the measured growth rate of the large ripples (Fig.~2a, Methods). 


\bibliographystyle{naturemag}

\begin{thebibliography}{10}
\expandafter\ifx\csname url\endcsname\relax
  \def\url#1{\texttt{#1}}\fi
\expandafter\ifx\csname urlprefix\endcsname\relax\def\urlprefix{URL }\fi
\providecommand{\bibinfo}[2]{#2}
\providecommand{\eprint}[2][]{\url{#2}}

\bibitem{hayes2018dunes}
\bibinfo{author}{Hayes, A.~G.}
\newblock \bibinfo{title}{Dunes across the solar system}.
\newblock \emph{\bibinfo{journal}{Science}} \textbf{\bibinfo{volume}{360}},
  \bibinfo{pages}{960--961} (\bibinfo{year}{2018}).

\bibitem{Bourkeetal10}
\bibinfo{author}{Bourke, M.~C.} \emph{et~al.}
\newblock \bibinfo{title}{Extraterrestrial dunes: An introduction to the
  special issue on planetary dune systems}.
\newblock \emph{\bibinfo{journal}{Geomorphology}}
  \textbf{\bibinfo{volume}{121}}, \bibinfo{pages}{1--14}
  (\bibinfo{year}{2010}).

\bibitem{Rasmussenetal15}
\bibinfo{author}{Rasmussen, K.~R.}, \bibinfo{author}{Valance, A.} \&
  \bibinfo{author}{Merrison, J.}
\newblock \bibinfo{title}{Laboratory studies of aeolian sediment transport
  processes on planetary surfaces}.
\newblock \emph{\bibinfo{journal}{Geomorphology}}
  \textbf{\bibinfo{volume}{244}}, \bibinfo{pages}{74--94}
  (\bibinfo{year}{2015}).

\bibitem{Diniegaetal17}
\bibinfo{author}{Diniega, S.} \emph{et~al.}
\newblock \bibinfo{title}{Our evolving understanding of aeolian bedforms, based
  on observation of dunes on different worlds}.
\newblock \emph{\bibinfo{journal}{Aeolian Research}}
  \textbf{\bibinfo{volume}{26}}, \bibinfo{pages}{5--27} (\bibinfo{year}{2017}).

\bibitem{Charruetal13}
\bibinfo{author}{Charru, F.}, \bibinfo{author}{Andreotti, B.} \&
  \bibinfo{author}{Claudin, P.}
\newblock \bibinfo{title}{Sand ripples and dunes}.
\newblock \emph{\bibinfo{journal}{Annual Review of Fluid Mechanics}}
  \textbf{\bibinfo{volume}{45}}, \bibinfo{pages}{469--493}
  (\bibinfo{year}{2013}).

\bibitem{Telferetal18}
\bibinfo{author}{Telfer, M.~W.} \emph{et~al.}
\newblock \bibinfo{title}{Dunes on {Pluto}}.
\newblock \emph{\bibinfo{journal}{Science}} \textbf{\bibinfo{volume}{360}},
  \bibinfo{pages}{992--997} (\bibinfo{year}{2018}).

\bibitem{Ewingetal17}
\bibinfo{author}{Ewing, R.~C.} \emph{et~al.}
\newblock \bibinfo{title}{Sedimentary processes of the {Bagnold Dunes}:
  Implications for the eolian rock record of {Mars}}.
\newblock \emph{\bibinfo{journal}{Journal of Geophysical Research: Planets}}
  \textbf{\bibinfo{volume}{122}}, \bibinfo{pages}{2544--2573}
  (\bibinfo{year}{2017}).

\bibitem{Bakeretal18b}
\bibinfo{author}{Baker, M.~M.} \emph{et~al.}
\newblock \bibinfo{title}{The {Bagnold Dunes} in southern summer: Active
  sediment transport on {Mars} observed by the {Curiosity} rover}.
\newblock \emph{\bibinfo{journal}{Geophysical Research Letters}}
  \textbf{\bibinfo{volume}{45}}, \bibinfo{pages}{8853--8863}
  (\bibinfo{year}{2018}).

\bibitem{Siminovichetal19}
\bibinfo{author}{Siminovich, A.} \emph{et~al.}
\newblock \bibinfo{title}{Numerical study of shear stress distribution over
  sand ripples under terrestrial and {Martian} conditions}.
\newblock \emph{\bibinfo{journal}{Journal of Geophysical Research: Planets}}
  \textbf{\bibinfo{volume}{124}}, \bibinfo{pages}{175--185}
  (\bibinfo{year}{2019}).

\bibitem{Sullivanetal20}
\bibinfo{author}{Sullivan, R.}, \bibinfo{author}{Kok, J.~F.},
  \bibinfo{author}{Katra, I.} \& \bibinfo{author}{Yizhaq, H.}
\newblock \bibinfo{title}{A broad continuum of aeolian impact ripple
  morphologies on {Mars} is enabled by low wind dynamic pressures}.
\newblock \emph{\bibinfo{journal}{Journal of Geophysical Research: Planets}}
  \textbf{\bibinfo{volume}{125}}, \bibinfo{pages}{e2020JE006485}
  (\bibinfo{year}{2020}).

\bibitem{lorenz2020martian}
\bibinfo{author}{Lorenz, R.~D.}
\newblock \bibinfo{title}{Martian ripples making a splash}.
\newblock \emph{\bibinfo{journal}{Journal of Geophysical Research: Planets}}
  \textbf{\bibinfo{volume}{125}}, \bibinfo{pages}{e2020JE006658}
  (\bibinfo{year}{2020}).

\bibitem{Lapotreetal16}
\bibinfo{author}{Lapotre, M. G.~A.} \emph{et~al.}
\newblock \bibinfo{title}{Large wind ripples on {Mars}: A record of atmospheric
  evolution}.
\newblock \emph{\bibinfo{journal}{Science}} \textbf{\bibinfo{volume}{353}},
  \bibinfo{pages}{55--58} (\bibinfo{year}{2016}).

\bibitem{Lapotreetal18}
\bibinfo{author}{Lapotre, M. G.~A.} \emph{et~al.}
\newblock \bibinfo{title}{Morphologic diversity of {Martian} ripples:
  Implications for large-ripple formation}.
\newblock \emph{\bibinfo{journal}{Geophysical Research Letters}}
  \textbf{\bibinfo{volume}{45}}, \bibinfo{pages}{10229--10239}
  (\bibinfo{year}{2018}).

\bibitem{Duranetal19}
\bibinfo{author}{{Dur\'an Vinent}, O.}, \bibinfo{author}{Andreotti, B.},
  \bibinfo{author}{Claudin, P.} \& \bibinfo{author}{Winter, C.}
\newblock \bibinfo{title}{A unified model of ripples and dunes in water and
  planetary environments}.
\newblock \emph{\bibinfo{journal}{Nature Geoscience}}
  \textbf{\bibinfo{volume}{12}}, \bibinfo{pages}{345--350}
  (\bibinfo{year}{2019}).

\bibitem{Lapotreetal21}
\bibinfo{author}{Lap{\^o}tre, M. G.~A.}, \bibinfo{author}{Ewing, R.~C.} \&
  \bibinfo{author}{Lamb, M.~P.}
\newblock \bibinfo{title}{An evolving understanding of enigmatic large ripples
  on {Mars}}.
\newblock \emph{\bibinfo{journal}{Journal of Geophysical Research: Planets}}
  \textbf{\bibinfo{volume}{126}}, \bibinfo{pages}{e2020JE006729}
  (\bibinfo{year}{2021}).

\bibitem{yizhaq2021turbulent}
\bibinfo{author}{Yizhaq, H.} \emph{et~al.}
\newblock \bibinfo{title}{Turbulent shear flow over large martian ripples}.
\newblock \emph{\bibinfo{journal}{Journal of Geophysical Research: Planets}}
  \textbf{\bibinfo{volume}{126}}, \bibinfo{pages}{e2020JE006515}
  (\bibinfo{year}{2021}).

\bibitem{rubanenko2022distinct}
\bibinfo{author}{Rubanenko, L.}, \bibinfo{author}{Lap{\^o}tre, M.~G.},
  \bibinfo{author}{Ewing, R.~C.}, \bibinfo{author}{Fenton, L.~K.} \&
  \bibinfo{author}{Gunn, A.}
\newblock \bibinfo{title}{A distinct ripple-formation regime on mars revealed
  by the morphometrics of barchan dunes}.
\newblock \emph{\bibinfo{journal}{Nature Communications}}
  \textbf{\bibinfo{volume}{13}}, \bibinfo{pages}{1--7} (\bibinfo{year}{2022}).

\bibitem{silvestro2016dune}
\bibinfo{author}{Silvestro, S.}, \bibinfo{author}{Vaz, D.},
  \bibinfo{author}{Yizhaq, H.} \& \bibinfo{author}{Esposito, F.}
\newblock \bibinfo{title}{Dune-like dynamic of martian aeolian large ripples}.
\newblock \emph{\bibinfo{journal}{Geophysical Research Letters}}
  \textbf{\bibinfo{volume}{43}}, \bibinfo{pages}{8384--8389}
  (\bibinfo{year}{2016}).

\bibitem{Bridgesetal12b}
\bibinfo{author}{Bridges, N.~T.} \emph{et~al.}
\newblock \bibinfo{title}{Planet-wide sand motion on {Mars}}.
\newblock \emph{\bibinfo{journal}{Geology}} \textbf{\bibinfo{volume}{40}},
  \bibinfo{pages}{31--34} (\bibinfo{year}{2012}).

\bibitem{balme2008transverse}
\bibinfo{author}{Balme, M.}, \bibinfo{author}{Berman, D.~C.},
  \bibinfo{author}{Bourke, M.~C.} \& \bibinfo{author}{Zimbelman, J.~R.}
\newblock \bibinfo{title}{Transverse aeolian ridges (tars) on mars}.
\newblock \emph{\bibinfo{journal}{Geomorphology}}
  \textbf{\bibinfo{volume}{101}}, \bibinfo{pages}{703--720}
  (\bibinfo{year}{2008}).

\bibitem{zimbelman2010transverse}
\bibinfo{author}{Zimbelman, J.~R.}
\newblock \bibinfo{title}{Transverse aeolian ridges on mars: First results from
  hirise images}.
\newblock \emph{\bibinfo{journal}{Geomorphology}}
  \textbf{\bibinfo{volume}{121}}, \bibinfo{pages}{22--29}
  (\bibinfo{year}{2010}).

\bibitem{hugenholtz2017morphology}
\bibinfo{author}{Hugenholtz, C.~H.}, \bibinfo{author}{Barchyn, T.~E.} \&
  \bibinfo{author}{Boulding, A.}
\newblock \bibinfo{title}{Morphology of transverse aeolian ridges (tars) on
  mars from a large sample: Further evidence of a megaripple origin?}
\newblock \emph{\bibinfo{journal}{Icarus}} \textbf{\bibinfo{volume}{286}},
  \bibinfo{pages}{193--201} (\bibinfo{year}{2017}).

\bibitem{Sullivanetal08}
\bibinfo{author}{Sullivan, R.} \emph{et~al.}
\newblock \bibinfo{title}{Wind-driven particle mobility on {Mars}: Insights
  from {Mars Exploration Rover} observations at {``El Dorado''} and
  surroundings at {Gusev Crater}}.
\newblock \emph{\bibinfo{journal}{Journal of Geophysical Research: Planets}}
  \textbf{\bibinfo{volume}{113}} (\bibinfo{year}{2008}).

\bibitem{jerolmack2006spatial}
\bibinfo{author}{Jerolmack, D.~J.}, \bibinfo{author}{Mohrig, D.},
  \bibinfo{author}{Grotzinger, J.~P.}, \bibinfo{author}{Fike, D.~A.} \&
  \bibinfo{author}{Watters, W.~A.}
\newblock \bibinfo{title}{Spatial grain size sorting in eolian ripples and
  estimation of wind conditions on planetary surfaces: Application to meridiani
  planum, mars}.
\newblock \emph{\bibinfo{journal}{Journal of Geophysical Research: Planets}}
  \textbf{\bibinfo{volume}{111}} (\bibinfo{year}{2006}).

\bibitem{weitz2018sand}
\bibinfo{author}{Weitz, C.~M.} \emph{et~al.}
\newblock \bibinfo{title}{Sand grain sizes and shapes in eolian bedforms at
  gale crater, mars}.
\newblock \emph{\bibinfo{journal}{Geophysical Research Letters}}
  \textbf{\bibinfo{volume}{45}}, \bibinfo{pages}{9471--9479}
  (\bibinfo{year}{2018}).

\bibitem{Bagnold41}
\bibinfo{author}{Bagnold, R.~A.}
\newblock \emph{\bibinfo{title}{The Physics of Blown Sand and Desert Dunes}}
  (\bibinfo{publisher}{Methuen, New York}, \bibinfo{year}{1941}).

\bibitem{southard1990bed}
\bibinfo{author}{Southard, J.~B.} \& \bibinfo{author}{Boguchwal, L.~A.}
\newblock \bibinfo{title}{Bed configuration in steady unidirectional water
  flows; part 2, synthesis of flume data}.
\newblock \emph{\bibinfo{journal}{Journal of Sedimentary Research}}
  \textbf{\bibinfo{volume}{60}}, \bibinfo{pages}{658--679}
  (\bibinfo{year}{1990}).

\bibitem{greeley1987wind}
\bibinfo{author}{Greeley, R.} \& \bibinfo{author}{Iversen, J.~D.}
\newblock \emph{\bibinfo{title}{Wind as a geological process: on Earth, Mars,
  Venus and Titan}}.
\newblock \bibinfo{number}{4} (\bibinfo{publisher}{CUP Archive},
  \bibinfo{year}{1987}).

\bibitem{Wilson72}
\bibinfo{author}{Wilson, I.~G.}
\newblock \bibinfo{title}{Aeolian bedforms---{Their} development and origins}.
\newblock \emph{\bibinfo{journal}{Sedimentology}}
  \textbf{\bibinfo{volume}{19}}, \bibinfo{pages}{173--210}
  (\bibinfo{year}{1972}).

\bibitem{abrams1985relaxation}
\bibinfo{author}{Abrams, J.} \& \bibinfo{author}{Hanratty, T.~J.}
\newblock \bibinfo{title}{Relaxation effects observed for turbulent flow over a
  wavy surface}.
\newblock \emph{\bibinfo{journal}{Journal of Fluid Mechanics}}
  \textbf{\bibinfo{volume}{151}}, \bibinfo{pages}{443--455}
  (\bibinfo{year}{1985}).

\bibitem{frederick1988velocity}
\bibinfo{author}{Frederick, K.~A.} \& \bibinfo{author}{Hanratty, T.~J.}
\newblock \bibinfo{title}{Velocity measurements for a turbulent nonseparated
  flow over solid waves}.
\newblock \emph{\bibinfo{journal}{Experiments in fluids}}
  \textbf{\bibinfo{volume}{6}}, \bibinfo{pages}{477--486}
  (\bibinfo{year}{1988}).

\bibitem{claudin2017dissolution}
\bibinfo{author}{Claudin, P.}, \bibinfo{author}{Dur{\'a}n, O.} \&
  \bibinfo{author}{Andreotti, B.}
\newblock \bibinfo{title}{Dissolution instability and roughening transition}.
\newblock \emph{\bibinfo{journal}{Journal of Fluid Mechanics}}
  \textbf{\bibinfo{volume}{832}} (\bibinfo{year}{2017}).

\bibitem{ellwood1975small}
\bibinfo{author}{Ellwood, J.~M.}, \bibinfo{author}{Evans, P.~D.} \&
  \bibinfo{author}{Wilson, I.~G.}
\newblock \bibinfo{title}{Small scale aeolian bedforms}.
\newblock \emph{\bibinfo{journal}{Journal of Sedimentary Research}}
  \textbf{\bibinfo{volume}{45}}, \bibinfo{pages}{554--561}
  (\bibinfo{year}{1975}).

\bibitem{Walker81}
\bibinfo{author}{Walker, J.~D.}
\newblock \emph{\bibinfo{title}{An experimental study on wind ripples, Master
  Thesis}}.
\newblock Master's thesis, \bibinfo{school}{Massachusetts Institute of
  Technology}, \bibinfo{address}{Cambridge, Massachusetts}
  (\bibinfo{year}{1981}).
\newblock \urlprefix\url{https://dspace.mit.edu/handle/1721.1/16156}.

\bibitem{Schmerleretal16}
\bibinfo{author}{Schmerler, E.}, \bibinfo{author}{Katra, I.},
  \bibinfo{author}{Kok, J.~F.}, \bibinfo{author}{Tsoar, H.} \&
  \bibinfo{author}{Yizhaq, H.}
\newblock \bibinfo{title}{Experimental and numerical study of sharp's shadow
  zone hypothesis on sand ripple wavelength}.
\newblock \emph{\bibinfo{journal}{Aeolian Research}}
  \textbf{\bibinfo{volume}{22}}, \bibinfo{pages}{37--46}
  (\bibinfo{year}{2016}).

\bibitem{Chengetal18}
\bibinfo{author}{Cheng, H.} \emph{et~al.}
\newblock \bibinfo{title}{Experimental study of aeolian sand ripples in a wind
  tunnel}.
\newblock \emph{\bibinfo{journal}{Earth Surface Processes and Landforms}}
  \textbf{\bibinfo{volume}{43}}, \bibinfo{pages}{312--321}
  (\bibinfo{year}{2018}).

\bibitem{Andreottietal06}
\bibinfo{author}{Andreotti, B.}, \bibinfo{author}{Claudin, P.} \&
  \bibinfo{author}{Pouliquen, O.}
\newblock \bibinfo{title}{Aeolian sand ripples: Experimental study of fully
  developed states}.
\newblock \emph{\bibinfo{journal}{Physical Review Letters}}
  \textbf{\bibinfo{volume}{96}}, \bibinfo{pages}{028001}
  (\bibinfo{year}{2006}).

\bibitem{Sharp63}
\bibinfo{author}{Sharp, R.~P.}
\newblock \bibinfo{title}{Wind ripples}.
\newblock \emph{\bibinfo{journal}{The Journal of Geology}}
  \textbf{\bibinfo{volume}{71}}, \bibinfo{pages}{617--636}
  (\bibinfo{year}{1963}).

\bibitem{cornish1897formation}
\bibinfo{author}{Cornish, V.}
\newblock \bibinfo{title}{On the formation of sand-dunes}.
\newblock \emph{\bibinfo{journal}{The Geographical Journal}}
  \textbf{\bibinfo{volume}{9}}, \bibinfo{pages}{278--302}
  (\bibinfo{year}{1897}).

\bibitem{SeppalaLinde78}
\bibinfo{author}{Sepp{\"a}l{\"a}, M.} \& \bibinfo{author}{Lind{\'e}, K.}
\newblock \bibinfo{title}{Wind tunnel studies of ripple formation}.
\newblock \emph{\bibinfo{journal}{Geografiska Annaler: Series A, Physical
  Geography}} \textbf{\bibinfo{volume}{60}}, \bibinfo{pages}{29--42}
  (\bibinfo{year}{1978}).

\bibitem{ling1998wind}
\bibinfo{author}{Ling, Y.}, \bibinfo{author}{Wu, Z.} \& \bibinfo{author}{Liu,
  S.}
\newblock \bibinfo{title}{A wind tunnel simulation of aeolian sand ripple
  formation}.
\newblock \emph{\bibinfo{journal}{ACTA GEOGRAPHICA SINICA-CHINESE EDITION-}}
  \textbf{\bibinfo{volume}{53}}, \bibinfo{pages}{527--534}
  (\bibinfo{year}{1998}).

\bibitem{miller1987wind}
\bibinfo{author}{Miller, J.}, \bibinfo{author}{Marshall, J.} \&
  \bibinfo{author}{Greeley, R.}
\newblock \bibinfo{title}{Wind ripples in low density atmospheres.}
\newblock \emph{\bibinfo{journal}{NASA Tech. Memo., NASA TM-89810}}
  \bibinfo{pages}{268--270} (\bibinfo{year}{1987}).

\bibitem{Duranetal14b}
\bibinfo{author}{Dur\'an, O.}, \bibinfo{author}{Claudin, P.} \&
  \bibinfo{author}{Andreotti, B.}
\newblock \bibinfo{title}{Direct numerical simulations of aeolian sand
  ripples}.
\newblock \emph{\bibinfo{journal}{Proceedings of the National Academy of
  Sciences of the United States of America}} \textbf{\bibinfo{volume}{111}},
  \bibinfo{pages}{15665--15668} (\bibinfo{year}{2014}).

\bibitem{Fourriereetal10}
\bibinfo{author}{Fourri\`ere, A.}, \bibinfo{author}{Claudin, P.} \&
  \bibinfo{author}{Andreotti, B.}
\newblock \bibinfo{title}{Bedforms in a turbulent stream: formation of ripples
  by primary linear instability and of dunes by nonlinear pattern coarsening}.
\newblock \emph{\bibinfo{journal}{Journal of Fluid Mechanics}}
  \textbf{\bibinfo{volume}{649}}, \bibinfo{pages}{287--328}
  (\bibinfo{year}{2010}).

\bibitem{ColombiniStocchino11}
\bibinfo{author}{Colombini, M.} \& \bibinfo{author}{Stocchino, A.}
\newblock \bibinfo{title}{Ripple and dune formation in rivers}.
\newblock \emph{\bibinfo{journal}{Journal of Fluid Mechanics}}
  \textbf{\bibinfo{volume}{673}}, \bibinfo{pages}{121--131}
  (\bibinfo{year}{2011}).

\bibitem{rubin1980single}
\bibinfo{author}{Rubin, D.} \& \bibinfo{author}{McCulloch, D.}
\newblock \bibinfo{title}{Single and superimposed bedforms: a synthesis of san
  francisco bay and flume observations}.
\newblock \emph{\bibinfo{journal}{Sedimentary Geology}}
  \textbf{\bibinfo{volume}{26}}, \bibinfo{pages}{207--231}
  (\bibinfo{year}{1980}).

\bibitem{lapotre2017sets}
\bibinfo{author}{Lapotre, M.~G.}, \bibinfo{author}{Lamb, M.~P.} \&
  \bibinfo{author}{McElroy, B.}
\newblock \bibinfo{title}{What sets the size of current ripples?}
\newblock \emph{\bibinfo{journal}{Geology}} \textbf{\bibinfo{volume}{45}},
  \bibinfo{pages}{243--246} (\bibinfo{year}{2017}).

\bibitem{grazer1982experimental}
\bibinfo{author}{Grazer, R.~A.}
\newblock \emph{\bibinfo{title}{Experimental study of current ripples using
  medium silt}}.
\newblock Ph.D. thesis, \bibinfo{school}{Massachusetts Institute of Technology}
  (\bibinfo{year}{1982}).

\bibitem{yalin1985determination}
\bibinfo{author}{Yalin, M.~S.}
\newblock \bibinfo{title}{On the determination of ripple geometry}.
\newblock \emph{\bibinfo{journal}{Journal of Hydraulic Engineering}}
  \textbf{\bibinfo{volume}{111}}, \bibinfo{pages}{1148--1155}
  (\bibinfo{year}{1985}).

\bibitem{Sauermannetal01}
\bibinfo{author}{Sauermann, G.}, \bibinfo{author}{Kroy, K.} \&
  \bibinfo{author}{Herrmann, H.~J.}
\newblock \bibinfo{title}{A continuum saltation model for sand dunes}.
\newblock \emph{\bibinfo{journal}{Physical Review E}}
  \textbf{\bibinfo{volume}{64}}, \bibinfo{pages}{031305}
  (\bibinfo{year}{2001}).

\bibitem{Duranetal11}
\bibinfo{author}{Dur\'an, O.}, \bibinfo{author}{Claudin, P.} \&
  \bibinfo{author}{Andreotti, B.}
\newblock \bibinfo{title}{On aeolian transport: {Grain-scale} interactions,
  dynamical mechanisms and scaling laws}.
\newblock \emph{\bibinfo{journal}{Aeolian Research}}
  \textbf{\bibinfo{volume}{3}}, \bibinfo{pages}{243--270}
  (\bibinfo{year}{2011}).

\bibitem{Elbelrhitietal05}
\bibinfo{author}{Elbelrhiti, H.}, \bibinfo{author}{Claudin, P.} \&
  \bibinfo{author}{Andreotti, B.}
\newblock \bibinfo{title}{Field evidence for surface-wave-induced instability
  of sand dunes}.
\newblock \emph{\bibinfo{journal}{Nature}} \textbf{\bibinfo{volume}{437}},
  \bibinfo{pages}{720--723} (\bibinfo{year}{2005}).

\bibitem{Andreottietal10}
\bibinfo{author}{Andreotti, B.}, \bibinfo{author}{Claudin, P.} \&
  \bibinfo{author}{Pouliquen, O.}
\newblock \bibinfo{title}{Measurements of the aeolian sand transport saturation
  length}.
\newblock \emph{\bibinfo{journal}{Geomorphology}}
  \textbf{\bibinfo{volume}{123}}, \bibinfo{pages}{343--348}
  (\bibinfo{year}{2010}).

\bibitem{Selmanietal18}
\bibinfo{author}{Selmani, H.}, \bibinfo{author}{Valance, A.},
  \bibinfo{author}{{Ould El Moctar}, A.}, \bibinfo{author}{Dupont, P.} \&
  \bibinfo{author}{Zegadi, R.}
\newblock \bibinfo{title}{Aeolian sand transport in out-of-equilibrium
  regimes}.
\newblock \emph{\bibinfo{journal}{Geophysical Research Letters}}
  \textbf{\bibinfo{volume}{45}}, \bibinfo{pages}{1838--1844}
  (\bibinfo{year}{2018}).

\bibitem{lu2021direct}
\bibinfo{author}{L{\"u}, P.} \emph{et~al.}
\newblock \bibinfo{title}{Direct validation of dune instability theory}.
\newblock \emph{\bibinfo{journal}{Proceedings of the National Academy of
  Sciences}} \textbf{\bibinfo{volume}{118}}, \bibinfo{pages}{e2024105118}
  (\bibinfo{year}{2021}).

\bibitem{Anderson87}
\bibinfo{author}{Anderson, R.~S.}
\newblock \bibinfo{title}{A theoretical model for aeolian impact ripples}.
\newblock \emph{\bibinfo{journal}{Sedimentology}}
  \textbf{\bibinfo{volume}{34}}, \bibinfo{pages}{943--956}
  (\bibinfo{year}{1987}).

\bibitem{Andreotti04}
\bibinfo{author}{Andreotti, B.}
\newblock \bibinfo{title}{A two-species model of aeolian sand transport}.
\newblock \emph{\bibinfo{journal}{Journal of Fluid Mechanics}}
  \textbf{\bibinfo{volume}{510}}, \bibinfo{pages}{47--70}
  (\bibinfo{year}{2004}).

\bibitem{Lammeletal12}
\bibinfo{author}{L\"ammel, M.}, \bibinfo{author}{Rings, D.} \&
  \bibinfo{author}{Kroy, K.}
\newblock \bibinfo{title}{A two-species continuum model for aeolian sand
  transport}.
\newblock \emph{\bibinfo{journal}{New Journal of Physics}}
  \textbf{\bibinfo{volume}{14}}, \bibinfo{pages}{093037}
  (\bibinfo{year}{2012}).

\bibitem{Lammeletal17}
\bibinfo{author}{L\"ammel, M.}, \bibinfo{author}{Dzikowski, K.},
  \bibinfo{author}{Kroy, K.}, \bibinfo{author}{Oger, L.} \&
  \bibinfo{author}{Valance, A.}
\newblock \bibinfo{title}{Grain-scale modeling and splash parametrization for
  aeolian sand transport}.
\newblock \emph{\bibinfo{journal}{Physical Review E}}
  \textbf{\bibinfo{volume}{95}}, \bibinfo{pages}{022902}
  (\bibinfo{year}{2017}).

\bibitem{tholen2023anomalous}
\bibinfo{author}{Tholen, K.}, \bibinfo{author}{P{\"a}htz, T.},
  \bibinfo{author}{Kamath, S.}, \bibinfo{author}{Parteli, E.~J.} \&
  \bibinfo{author}{Kroy, K.}
\newblock \bibinfo{title}{Anomalous scaling of aeolian sand transport reveals
  coupling to bed rheology}.
\newblock \emph{\bibinfo{journal}{Physical Review Letters}}
  \textbf{\bibinfo{volume}{130}}, \bibinfo{pages}{058204}
  (\bibinfo{year}{2023}).

\bibitem{AndersonBunas93}
\bibinfo{author}{Anderson, R.~S.} \& \bibinfo{author}{Bunas, K.~L.}
\newblock \bibinfo{title}{Grain size segregation and stratigraphy in aeolian
  ripples modelled with a cellular automaton}.
\newblock \emph{\bibinfo{journal}{Nature}} \textbf{\bibinfo{volume}{365}},
  \bibinfo{pages}{740} (\bibinfo{year}{1993}).

\bibitem{Lammeletal18}
\bibinfo{author}{L\"ammel, M.} \emph{et~al.}
\newblock \bibinfo{title}{Aeolian sand sorting and megaripple formation}.
\newblock \emph{\bibinfo{journal}{Nature Physics}}
  \textbf{\bibinfo{volume}{14}}, \bibinfo{pages}{759--765}
  (\bibinfo{year}{2018}).

\bibitem{tholen2022megaripple}
\bibinfo{author}{Tholen, K.}, \bibinfo{author}{P{\"a}htz, T.},
  \bibinfo{author}{Yizhaq, H.}, \bibinfo{author}{Katra, I.} \&
  \bibinfo{author}{Kroy, K.}
\newblock \bibinfo{title}{Megaripple mechanics: bimodal transport ingrained in
  bimodal sands}.
\newblock \emph{\bibinfo{journal}{Nature communications}}
  \textbf{\bibinfo{volume}{13}}, \bibinfo{pages}{1--11} (\bibinfo{year}{2022}).

\bibitem{Yizhaq04}
\bibinfo{author}{Yizhaq, H.}
\newblock \bibinfo{title}{A simple model of aeolian megaripples}.
\newblock \emph{\bibinfo{journal}{Physica A: Statistical Mechanics and its
  Applications}} \textbf{\bibinfo{volume}{338}}, \bibinfo{pages}{211--217}
  (\bibinfo{year}{2004}).

\bibitem{Makse00}
\bibinfo{author}{Makse, H.~A.}
\newblock \bibinfo{title}{Grain segregation mechanism in aeolian sand ripples}.
\newblock \emph{\bibinfo{journal}{The European Physical Journal E}}
  \textbf{\bibinfo{volume}{1}}, \bibinfo{pages}{127--135}
  (\bibinfo{year}{2000}).

\bibitem{ManukyanPrigozhin09}
\bibinfo{author}{Manukyan, E.} \& \bibinfo{author}{Prigozhin, L.}
\newblock \bibinfo{title}{Formation of aeolian ripples and sand sorting}.
\newblock \emph{\bibinfo{journal}{Physical Review E}}
  \textbf{\bibinfo{volume}{79}}, \bibinfo{pages}{031303}
  (\bibinfo{year}{2009}).

\bibitem{pye2008aeolian}
\bibinfo{author}{Pye, K.} \& \bibinfo{author}{Tsoar, H.}
\newblock \emph{\bibinfo{title}{Aeolian sand and sand dunes}}
  (\bibinfo{publisher}{Springer Science \& Business Media},
  \bibinfo{year}{2008}).

\bibitem{lancaster2013geomorphology}
\bibinfo{author}{Lancaster, N.}
\newblock \emph{\bibinfo{title}{Geomorphology of desert dunes}}
  (\bibinfo{publisher}{Routledge}, \bibinfo{year}{2013}).

\bibitem{Yizhaqetal09}
\bibinfo{author}{Yizhaq, H.}, \bibinfo{author}{Isenberg, O.},
  \bibinfo{author}{Wenkart, R.}, \bibinfo{author}{Tsoar, H.} \&
  \bibinfo{author}{Karnieli, A.}
\newblock \bibinfo{title}{Morphology and dynamics of aeolian mega-ripples in
  {Nahal Kasuy}, southern {Israel}}.
\newblock \emph{\bibinfo{journal}{Israel Journal of Earth Sciences}}
  \textbf{\bibinfo{volume}{57}}, \bibinfo{pages}{149--165}
  (\bibinfo{year}{2009}).

\bibitem{KatraYizhaq17}
\bibinfo{author}{Katra, I.} \& \bibinfo{author}{Yizhaq, H.}
\newblock \bibinfo{title}{Intensity and degree of segregation in bimodal and
  multimodal grain size distributions}.
\newblock \emph{\bibinfo{journal}{Aeolian Research}}
  \textbf{\bibinfo{volume}{27}}, \bibinfo{pages}{23--34}
  (\bibinfo{year}{2017}).

\bibitem{McKennaNeumanBedard17}
\bibinfo{author}{{McKenna Neuman}, C.} \& \bibinfo{author}{B\'edard, O.}
\newblock \bibinfo{title}{A wind tunnel investigation of particle segregation,
  ripple formation and armouring within sand beds of systematically varied
  texture}.
\newblock \emph{\bibinfo{journal}{Earth Surface Processes and Landforms}}
  \textbf{\bibinfo{volume}{42}}, \bibinfo{pages}{749--762}
  (\bibinfo{year}{2017}).

\bibitem{yizhaq2012a}
\bibinfo{author}{Yizhaq, H.}, \bibinfo{author}{Katra, I.},
  \bibinfo{author}{Kok, J.~F.} \& \bibinfo{author}{Isenberg, O.}
\newblock \bibinfo{title}{Transverse instability of megaripples}.
\newblock \emph{\bibinfo{journal}{Geology}} \textbf{\bibinfo{volume}{40}},
  \bibinfo{pages}{459--462} (\bibinfo{year}{2012}).

\bibitem{PyeTsoar09}
\bibinfo{author}{Pye, K.} \& \bibinfo{author}{Tsoar, H.}
\newblock \emph{\bibinfo{title}{Aeolian Sand and Sand Dunes}}
  (\bibinfo{publisher}{Springer, Berlin}, \bibinfo{year}{2009}).

\bibitem{Katraetal14}
\bibinfo{author}{Katra, I.}, \bibinfo{author}{H, Y.} \& \bibinfo{author}{Kok,
  J.~F.}
\newblock \bibinfo{title}{Mechanisms limiting the growth of aeolian
  megaripples}.
\newblock \emph{\bibinfo{journal}{Geophysical Research Letters}}
  \textbf{\bibinfo{volume}{41}}, \bibinfo{pages}{858--865}
  (\bibinfo{year}{2014}).

\bibitem{Andreottietal21}
\bibinfo{author}{Andreotti, B.}, \bibinfo{author}{Claudin, P.},
  \bibinfo{author}{Iversen, J.~J.}, \bibinfo{author}{Merrison, J.~P.} \&
  \bibinfo{author}{Rasmussen, K.~R.}
\newblock \bibinfo{title}{A lower than expected saltation threshold at
  {Martian} pressure and below}.
\newblock \emph{\bibinfo{journal}{Proceedings of the National Academy of
  Sciences of the United States of America}} \textbf{\bibinfo{volume}{118}},
  \bibinfo{pages}{e2012386118} (\bibinfo{year}{2021}).

\bibitem{holstein2014environmental}
\bibinfo{author}{Holstein-Rathlou, C.} \emph{et~al.}
\newblock \bibinfo{title}{An environmental wind tunnel facility for testing
  meteorological sensor systems}.
\newblock \emph{\bibinfo{journal}{Journal of atmospheric and oceanic
  technology}} \textbf{\bibinfo{volume}{31}}, \bibinfo{pages}{447--457}
  (\bibinfo{year}{2014}).

\bibitem{pahtz2023scaling}
\bibinfo{author}{P{\"a}htz, T.} \& \bibinfo{author}{Dur{\'a}n, O.}
\newblock \bibinfo{title}{Scaling laws for planetary sediment transport from
  dem-rans numerical simulations}.
\newblock \emph{\bibinfo{journal}{Journal of Fluid Mechanics}}
  \textbf{\bibinfo{volume}{963}}, \bibinfo{pages}{A20} (\bibinfo{year}{2023}).

\bibitem{Andreottietal09}
\bibinfo{author}{Andreotti, B.}, \bibinfo{author}{Fourri\`ere, A.},
  \bibinfo{author}{Ould-Kaddour, F.}, \bibinfo{author}{Murray, B.} \&
  \bibinfo{author}{Claudin, P.}
\newblock \bibinfo{title}{Giant aeolian dune size determined by the average
  depth of the atmospheric boundary layer}.
\newblock \emph{\bibinfo{journal}{Nature}} \textbf{\bibinfo{volume}{457}},
  \bibinfo{pages}{1120--1123} (\bibinfo{year}{2009}).

\bibitem{Kroyetal02a}
\bibinfo{author}{Kroy, K.}, \bibinfo{author}{Sauermann, G.} \&
  \bibinfo{author}{Herrmann, H.~J.}
\newblock \bibinfo{title}{Minimal model for aeolian sand dunes}.
\newblock \emph{\bibinfo{journal}{Physical Review E}}
  \textbf{\bibinfo{volume}{66}}, \bibinfo{pages}{031302}
  (\bibinfo{year}{2002}).

\bibitem{Kroyetal02b}
\bibinfo{author}{Kroy, K.}, \bibinfo{author}{Sauermann, G.} \&
  \bibinfo{author}{Herrmann, H.~J.}
\newblock \bibinfo{title}{A minimal model for sand dunes}.
\newblock \emph{\bibinfo{journal}{Physical Review Letters}}
  \textbf{\bibinfo{volume}{64}}, \bibinfo{pages}{054301}
  (\bibinfo{year}{2002}).

\bibitem{Andreottietal02b}
\bibinfo{author}{Andreotti, B.}, \bibinfo{author}{Claudin, P.} \&
  \bibinfo{author}{Douady, S.}
\newblock \bibinfo{title}{Selection of dune shapes and velocities part 2: A
  two-dimensional modelling}.
\newblock \emph{\bibinfo{journal}{The European Physical Journal B}}
  \textbf{\bibinfo{volume}{28}}, \bibinfo{pages}{341--352}
  (\bibinfo{year}{2002}).

\bibitem{miller1977threshold}
\bibinfo{author}{Miller, M.}, \bibinfo{author}{McCave, I.} \&
  \bibinfo{author}{Komar, P.}
\newblock \bibinfo{title}{Threshold of sediment motion under unidirectional
  currents}.
\newblock \emph{\bibinfo{journal}{Sedimentology}}
  \textbf{\bibinfo{volume}{24}}, \bibinfo{pages}{507--527}
  (\bibinfo{year}{1977}).

\bibitem{greeley1974wind}
\bibinfo{author}{Greeley, R.}, \bibinfo{author}{Iversen, J.},
  \bibinfo{author}{Pollack, J.}, \bibinfo{author}{Udovich, N.} \&
  \bibinfo{author}{White, B.}
\newblock \bibinfo{title}{Wind tunnel studies of martian aeolian processes}.
\newblock \emph{\bibinfo{journal}{Proceedings of the Royal Society of London.
  A. Mathematical and Physical Sciences}} \textbf{\bibinfo{volume}{341}},
  \bibinfo{pages}{331--360} (\bibinfo{year}{1974}).

\bibitem{JacksonHunt75}
\bibinfo{author}{Jackson, P.~S.} \& \bibinfo{author}{Hunt, J. C.~R.}
\newblock \bibinfo{title}{Turbulent wind flow over a low hill}.
\newblock \emph{\bibinfo{journal}{Quarterly Journal of the Royal Meteorological
  Society}} \textbf{\bibinfo{volume}{101}}, \bibinfo{pages}{929--955}
  (\bibinfo{year}{1975}).

\end{thebibliography}
 \newcommand{\noop}[1]{}



\setcounter{equation}{0}
\setcounter{table}{0}
\setcounter{section}{0}
\renewcommand{\theequation}{M\arabic{equation}}
\renewcommand{\thetable}{M\arabic{table}}
\renewcommand{\thesection}{M\arabic{section}}

\newpage
\section*{METHODS}

\section{Ambient-air wind tunnel}
The ambient-air experiments were conducted at the stationary boundary layer wind tunnel of the Aeolian Simulation Laboratory in Ben-Gurion University described in Refs.\cite{PyeTsoar09,Katraetal14} (Fig.~S1).  The test section of the tunnel is 7m long and has a square cross-section of $0.7 \times 0.7 ~\mathrm{m}$. We use artificial spherical grains (glass beads) of diameter $90 \pm 35~\mathrm{\mu m}$ and density $\rho_p = 2500~\mathrm{kg/m^3}$ (Fig.~S2). A constant wind was imposed on an initially $3~\mathrm{cm}$-thick sand layer for $80 ~\mathrm{min}$, while sand was fed upwind at a constant rate to minimize net bed erosion. Bed elevation profiles $h(x,t)$ were measured every minute along a $60~\mathrm{cm}$ central transect, using an inclined laser sheet coupled with a high-speed camera (Fig.~S1) with vertical and horizontal resolution $\Delta h = 0.5~\mathrm{mm}$ and $\Delta x=0.3~\mathrm{mm}$, respectively. To better resolve the very-small amplitude, centimeter-scale ripples, top-view photos were taken every $20$ seconds using a second camera with a resolution of $\Delta x=0.18~\mathrm{mm}$. Wind velocity was measured at different heights using Pitot-static tubes at the downwind end of the test section. Measured and estimated quantities are summarized in Supplementary Table~S1.


\section{Low-pressure wind tunnel}
The low-pressure experiments were conducted in a pressure-controlled wind tunnel of the Aarhus University  (AWTSII), described in Ref.\cite{Andreottietal21, holstein2014environmental}.  The test section of the tunnel is $3.25~\mathrm{m}$ long and has a rectangular cross section about $0.36~\mathrm{m}$ wide and $0.85~\mathrm{m}$ high. Experiments were run using the same sand as in the ambient-air experiments (Fig.~S2) for different particle-fluid density ratios $\rho_p/\rho_f$ by adjusting the interior pressure in the hermetically closed container with an uncertainty of $2\%$. For a given fluid pressure $P$ and constant temperature, the fluid density $\rho_f$ and kinematic viscosity $\nu$ relative to ambient-air conditions ($P_a=1000~\mathrm{mb}$, $\rho_{fa}=1.2~\mathrm{kg/m^3}$ and $\nu_a = 1.5\times10^{-5}~\mathrm{m^2/s}$) are estimated from the relation $\nu/\nu_a  = \rho_{fa}/\rho_f = P_a/P$. A single top-view photo was taken after about $1-2~\mathrm{h}$ to estimate the size of the mature bedforms (Fig. ~S3). When the ripple experiments were finished, the bed was flattened and wind velocity profiles were measured using Pitot-static tubes at different elevations. Measured and estimated quantities are summarized in Table~S2. 


\section{Impact-ripple numerical simulation}\label{simulation}
The formation of impact ripples was simulated using a grain-based sediment transport model as in Ref.\cite{Duranetal14b}. Simulations were performed for conditions similar to the wind tunnel experiments, as summarized in Table~S3 and shown in Fig.~1f.  In particular,  one parameter set represents the environmental conditions of the ambient-air wind tunnel experiments to provide a 1-to-1 comparison (see lower dashed line in Fig. 2b).  Bed elevation profiles $h(x,t)$ are calculated from mass conservation following Ref.\cite{Duranetal14b}.  An exemplary spatio-temporal evolution of ripples for ambient-air conditions is shown in Fig.~S4.


\section{Data extraction}

\subsection{Estimation of wind friction velocity.}
In both the ambient-air and low-pressure wind tunnel data, the velocity profile can be well approximated by the standard log-profile $U(z) = u_\ast/\kappa \ln{(z/z_0)}$ for a hydrodynamically-smooth regime with roughness length $z_0 = 0.11 \nu/u_\ast$ (Fig.~S5), where $\kappa = 0.4$ and $\nu$ is the kinematic viscosity. For all cases, the wind friction velocities $u_\ast$, estimated from fitting the velocity profile (Figs.~S5 and S6), are relatively close to the predicted transport threshold\cite{pahtz2023scaling}. In particular, for the low-pressure wind tunnel data, the variable friction velocity closely corresponds to a constant Shields number $\Theta = u_\ast^2/[(\rho_p/\rho_f-1)g d] = 0.0117 \pm 10^{-6}$ (Fig.~S6).  

\subsection{Estimation of bedform wavelengths from top-view photos.}
In both the ambient-air and low-pressure wind tunnel experiments, the wavelength of small and large ripples was obtained from top-view photographs using the autocorrelation function $C(\ell)$ of the normalized average $\eta(x)$ of the RGB channels over an arbitrary transect along the wind direction (Figs.~S7 and S8)
\begin{equation}
C(\ell)=\int \eta(x)\eta(x+\ell) \d x,
\end{equation}
where $\ell$ is the lag between the profiles. When two distinct bedforms are present, the wavelengths $\lambda_{\rm IR}$ and $\lambda_{\rm HR}$ of the smaller and larger bedform, respectively, correspond to the first local maximum and the  absolute maximum after the negative minimum of $C(\ell)$ as shown in Figs.~S7 and S8\cite{Andreottietal09}. When the first local maximum was not clearly defined, we extracted a portion of the profile and detrended it by subtracting a quadratic polynomial such that only the small bedform signal remained (Figs.~S8D). Finally, with only one bedform present, both maxima coincide (e.g. Figs.~S8A and B). 
We evaluated the uncertainty in bedform wavelength by using different transects covering most of the recorded field of view. 

\subsection{Estimation of bedform wavelength, amplitude and growth rate from ambient-air bed elevation profiles.}
The vertical resolution of the bed elevation profiles, measured using the inclined laser sheet, was not high enough to resolve the smaller ripples. Therefore, we numerically smoothened the measured profiles using a moving average over a $1~\mathrm{cm}$ window to reduce the noise. We also detrended the resulting profiles by subtracting a quadratic polynomial such that only the bedform signal remained. We then estimated the wavelength and amplitude of the bedform from the autocorrelation function $C(\ell;t)$ of the time series of the resulting elevation profiles $h(x,t)$,
\begin{equation}
C(\ell;t)=\int h(x,t)h(x+\ell,t) \d x.
\end{equation}
The ripple wavelength $\lambda(t)$ was determined as the location of the first maximum after the negative minimum, with the corresponding amplitude $A(t)=\sqrt{2C(\lambda(t))}$, as shown in Fig.~S9\cite{Andreottietal06,Duranetal19}. At early times ($t\lesssim 10 ~\mathrm{min}$), the inferred wavelengths and amplitudes of the large ripples in the  ambient-air wind tunnel are noisy because of the superimposed small ripples (Fig.~2). Hence, the growth rate of large ripples (shown in Fig.~4) was estimated by fitting the whole time series $A(t)$ by a hyperbolic tangent function (dotted line in Fig.~2a),
\begin{equation}
A(t)\propto \tanh(\sigma (t-t_0)),
\end{equation}
which reduces to an exponential function with growth rate $\sigma = 0.180 \pm 0.005~\mathrm{min^{-1}}$ at early times (dashed line in Fig.~2a). A similar method was used to estimate the initial wavelength of simulated ripples, summarized in Table~S3. 


\subsection{Estimation of the subscale saturation length.}

Following the definition of the saturation length as the spatial relaxation length of the sand flux (Supplementary Methods)\cite{Kroyetal02a,Kroyetal02b, Andreottietal02b,Andreottietal10}, the subscale saturation length $\ell_{\rm sat}^{\rm sub}$ can be estimated as the location of the maximum of a modified cross-correlation function between the sand flux perturbation $\delta q$ and the bed shear stress perturbation $\delta \tau$ (Fig.~4):
\begin{equation}
  C_{\delta q,\delta \tau}(\ell;t) = \int \delta q(x-\ell,t) \delta\tau(x,t) G(\delta q, \delta \tau) \d x  ,
\end{equation}
where the auxiliary function $G = \Theta(\delta q)\Theta(\delta \tau)/ \sqrt{\langle\delta q^2 \rangle \langle\delta \tau^2 \rangle}$, with $ \langle \cdot \rangle$ the spatial average and $\Theta(x)$ the Heaviside step function,  ensures that the cross-correlation is normalized and limited to positive perturbations ($\delta q > 0$ and $\delta \tau > 0$), related to ripple crests.

We first obtained the spatial variation of the sand flux $q(x)$ from the time evolution of the smoothed and detrended bed elevation profiles $h(x,t)$, acquired every $\Delta t = 1~\mathrm{min}$ in the ambient-air wind tunnel, by using mass conservation for bedforms with lateral uniformity ($\partial_y = 0$):
\begin{equation}\label{mass_conservation}
\partial_t h(x,t)= - \phi_b^{-1} \partial_x q(x) = \Gamma(x,t),
\end{equation}
where $\phi_b\approx 0.6$ is the bed volume fraction. We estimated the rate of elevation change $\Gamma$ as
\begin{equation}
  \Gamma(x,t) = \frac{\partial h}{\partial t} \approx \frac{h(x,t+\Delta t) - h(x,t)}{\Delta t}.
\end{equation}
Hence, the sand flux pertubation $ \delta q$ beyond a reference value $q(0,t)$ is given by
\begin{equation}
  \delta q(x,t) = q(x,t) - q(0,t) = - \phi_b \int_0^x \Gamma(x,t) dx.
\end{equation}
For the estimation of the perturbation of the normalized bed shear stress 
\begin{equation}\label{shearstress_pertubation}
  \delta \tau(x,t) = \tau(x,t)/\tau_0 -1 = f(\overline h, \mathcal{R}_p),
\end{equation}
due to the presence of bedforms, we used an existing hydrodynamic model\cite{Charruetal13,claudin2017dissolution,Duranetal19}, represented by the function $f$, that depends on the particle Reynolds number $\mathcal{R}_p = u_\ast d/ \nu$. Here,  $\tau_0$ is the undisturbed bed shear stress and $\overline h(x,t)$ is the average filtered profile $\overline h(x,t) = [h(x,t+\Delta t) + h(x,t)]/2 $.

\section{Comparison with water and Mars data}

In Fig.~1f, we show data of incipient subaqueous ripples with a saturation length above $100 \nu /u_\ast$, consistent with the defined wavelength corridor of hydrodynamic ripples, and particle Reynolds number less than $25$, i.e., excluding the hydrodyanmically rough regime (references are detailed in Supplementary Figs. 11–16 of Ref.\cite{Duranetal19}).  Data for Mars\cite{Lapotreetal16} as well as their corresponding environmental parameters are summarized in Supplementary Tabs.~1, 4–7 of Ref.\cite{Duranetal19}. The main sources of errors are also discussed in the Supplementary Information of Ref.\cite{Duranetal19}. Aeolian literature data near ambient-air conditions clearly distinguishing between both ripples could only be found in one historical study (Ref.\cite{miller1987wind}).  We estimated the corresponding wind friction velocities $u_\ast$ from their relation to the aeolian transport threshold using Miller's initiation threshold prediction $u_{it} \approx \sqrt{0.017 (\rho_p/\rho_f-1) gd}$ from a previous study (see Fig. 8 in Ref.\cite{miller1977threshold}), which is largely based on experiments using the same pressure-controlled wind tunnel\cite{greeley1974wind}.  As no errors were reported in these experiments, we use $\Delta\lambda=1 ~\mathrm{cm}$, accounting for the symbol size in Fig. 2  in Ref.\cite{miller1987wind}. Note that the resulting huge uncertainty of values is critical for the estimation of the wavelength corridor of the impact ripples (orange-shaded area in Fig.~ 1f), which therefore only serves to guide the eye. The ambient-air and low-pressure wind tunnel data points represent the average and standard deviation of all wavelength and viscous-sublayer-scale measurements,  respectively,  as described above (Tables~S1 and S2). 

\section{Wavelength corridor of mesoscale bedforms }

The wavelength corridor of hydrodynamic ripples (Fig.~1f, blue shaded area) is given by the condition $500 \nu/u_\ast < \lambda < 5000 \nu / u_\ast$. The lower value roughly characterizes the beginning of the transitional hydrodynamic response of the flow to the bed topography, where most hydrodynamic bedforms are expected\cite{Duranetal19}, whereas the upper value is the bound set by the hydrodynamic anomaly for the initial wavelength of hydrodynamic ripples, as predicted by the hydrodynamic bedform model\cite{Duranetal19} (cf. Fig. 2b in Ref.\cite{Duranetal19}).  Overall, the upper bound is consistent with the value $4000 \nu / u_\ast$, reported for subaqueous ripples\cite{lapotre2017sets}. The ``impact ripple'' corridor   (Fig.~1f, orange shaded area) only serves to guide the eye.


\section{New hydrodynamic bedform model}

This model represents an extension of the model in Ref.\cite{Duranetal19} accounting for the additional subscale relaxation process that we have postulated above.

\subsection{Transport relaxation and steady state.}
Following the postulated existence of a subscale saturation length $\lss$ in addition to the standard saturation length $\ls$ (characterizing the total spatial relaxation of the whole transport layer), we assume that a fraction $q^{\rm sub}$ of the total volumetric sand flux $q$ spatially relaxes towards the equilibrium saturated flux $q^{\rm sub}_s$ over a length $\lss$, whereas the remaining fraction $q^{\rm sup} = q - q^{\rm sub}$ relaxes towards the equilibrium saturated flux $q^{\rm sup}_s = q_s - q^{\rm sub}_s$ on the much larger ``super'' scale $\ls^{\rm sup}$. We also take the fraction of the total saturated flux $q_s$ that experiences a subscale modulation to be a constant $\gamma \equiv q^{\rm sub}_s / q_s$. Therefore, the spatial relaxation of the sand flux under nearly saturated conditions is described by the pair of linear equations (Eq. S1):
\begin{eqnarray}
  \partial_x q^{\rm sub} & = & (q^{\rm sub}_s - q^{\rm sub})/\lss, \\
  \partial_x q^{\rm sup} & = & (q^{\rm sup}_s - q^{\rm sup})/\ls^{\rm sup},
\end{eqnarray}
where $q^{\rm sub}_s = \gamma q_s$ and $q^{\rm sup}_s = (1-\gamma) q_s$ and, for simplicity, we identifiy $\ls^{\rm sup}$ with the measured (coarse-grained) saturation length $\ls$. For aeolian transport, the saturated flux $q_s$ is proportional to the bed shear stress $\tau$\cite{Duranetal11},
\begin{equation}
  q_s = Q_0 (\tau/\tau_t - 1),
\end{equation}
where, for fixed $\rho_p/\rho_f$\cite{pahtz2023scaling},  $Q_0$ is a dimensional constant and $\tau_t$ is the bed shear stress at the transport threshold.

\subsection{Linear instability analysis and dispersion relation.}

A low-amplitude (complex) surface perturbation $Z(x,t) = \epsilon e^{\Omega t + ik x}$ of wavelength $\lambda$ and wavenumber $k = 2 \pi/\lambda$ that is evolving with a complex growth rate $\Omega = \sigma - i c k$ will induce a modulation on both the bed shear stress ($\hat \tau \equiv \tau-\tau_0$) and the sand flux ($\hat q \equiv q -q_0$) proportional to $Z$, where $\tau_0 = \rho_f u_\ast^2$ and $q_0(\tau_0)$ are the undisturbed bed shear stress and saturated flux, respectively. Applying mass conservation (Eq. \ref{mass_conservation}) we get
\begin{equation}
  \Omega \, Z(x,t) = - \phi_b^{-1} ik \, \hat q(x,t),
\end{equation}
with $\hat q = \hat q^{\rm sub} + \hat q^{\rm sup}$. The modulated fluxes can be obtained from the relaxation equations (M9) and (M10) as
\begin{eqnarray}
  \hat q^{\rm sub} & = & \gamma \hat q_s/(1 + ik\lss), \\
  \hat q^{\rm sup} & = & (1-\gamma) \hat q_s/(1 + ik\ls) ,
\end{eqnarray}
where the modulated saturated flux is $\hat q_s = \partial_{\tau} q_s \hat \tau = \tau_t^{-1} Q_0 \hat \tau$. Finally, the modulated bed shear stress has the form\cite{JacksonHunt75} $\hat \tau(x,t) = \tau_0 (\mathcal{A} + i \mathcal{B}) Z(x,t)$, where $\mathcal{A}$ and $\mathcal{B}$ are the in-phase and in-quadrature components of the bed shear stress modulation, which are functions of the rescaled wavenumber $k \nu/u_\ast$ and the particle Reynolds number $\mathcal{R}_p$, analogeous to Eq. (\ref{shearstress_pertubation}). For conditions outside the hydrodynamically smooth regime, the particle Reynolds number is corrected for the transport feedback on the flow\cite{Charruetal13,Duranetal19}: 
\begin{equation}
 \mathcal{R}_p = u_\ast d \, e^{7 (1-u_t/u_\ast)}/\nu \,.
 \label{Rep}
\end{equation}
The dispersion relation $\sigma(k)$ is obtained by substituting all the terms back into the mass conservation (Eq. \ref{mass_conservation}) and taking its real part,
\begin{equation}
  \sigma^+(k) = (k d)^2 \left[ \gamma \left( \frac{\mathcal{B} - k \lss \mathcal{A}}{1+(k \lss)^2} \right) + (1-\gamma) \left( \frac{\mathcal{B} - k \ls \mathcal{A}}{1+(k \ls)^2} \right) \right],
  \label{Eq.disp_rel}
\end{equation}
where $\sigma^+$ is the dimensionless growth rate $\sigma^+ = \sigma\, \phi_b d^2 \tau_t \tau_0^{-1} Q_0^{-1}$, or, in terms of the saturated mass flux over a flat surface, $q_0$, the undisturbed friction velocity $u_\ast$ and the threshold friction velocity $u_t$: 
\begin{equation}
  \sigma^+ = \sigma\, \phi_b d^2 \rho_p [1-(u_t/u_\ast)^2]/ q_0.
  \label{sigma+}
\end{equation}
Figure~4 shows the dispersion relation for three cases: as if (i) only the standard saturation length ($\gamma = 0$) and only dunes existed (dotted line), (ii) only the subscale saturation length ($\gamma = 1$) and the associated hydrodynamic ripple mode existed, independent of a dune mode separated by a forbidden gap, as previously explained in terms of a hydrodynamic anomaly preventing bedform growth\cite{Duranetal19} (dashed line), and (iii) the complete model (solid line), where the wavelength gap expands as a result of the suppressing effect of the larger saturation process onto small wavelengths.

\subsection{Estimation of model parameters.}

The dispersion relation has two main unknown parameters: the weight $\gamma$ of the subscale modulation and its characteristic length scale, the subscale saturation length $\lss$. In a first approximation, we identified $\lss$ by the average hop length $\bar{\ell}_{\rm hop}$ of the complete transport layer, estimated as $7.7\pm0.3~\mathrm{mm}$ from grain-scale numerical simulations (Supplementary Methods). We then estimated $\gamma$ by fitting the dimensionless growth rate $\max(\sigma^+)$ of the fastest growing hydrodynamic ripple to the measured growth rate $0.18~\mathrm{min^{-1}}$, rescaled as in Eq.~(\ref{sigma+}), by using the ambient-air wind tunnel data (Table~S1) complemented with our estimation of the volumetric transport rate $q_0/\rho_p$ and threshold friction velocity $u_t$ from a recent transport model\cite{pahtz2023scaling} that is consistent with grain-scale numerical simulations  (Supplementary Methods). We find $\gamma = 0.1$, which means that only about $10\%$ of the sand flux can adapt locally, on the scale $\lss$.  Reassuringly, this small weight is consistent with a inconspicous, subdominant contribution of the postulated subscale process to the total sand ﬂux, and also justifies our above estimate of $\lss\approx\bar{\ell}_{\rm hop}$.

\subsection{Model validation.}

The dispersion relation Eq.~\ref{Eq.disp_rel} predicts that the fastest growing wavelength scales as $\lambda \approx 50 (\lss u_\ast/\nu)^{0.63} \nu/u_\ast$ (Fig.~S10), with the same exponent as expected from scaling arguments\cite{Lapotreetal16,lapotre2017sets,Lapotreetal21, rubanenko2022distinct}. It gives $\lambda = 7.1\pm 0.1 {\rm cm}$ for the ambient-air wind tunnel conditions, in good agreement with experimental data, as shown in Figs.~2b and 4. In Fig.~4, the experimental data point is the average over the first $15$ minutes of the (less noisy) wavelength values extracted from the top-view photos to  characterize the initial bedforms. Model predictions also compare well with measured wavelength and measured $\lss$ at different times, within the linear instability regime (Fig.~S10). 

To characterize the wind-tunnel conditions, we used the standard particle Reynolds number ($\mathcal{R}_p = u_\ast d/\nu$) instead of the corrected one (Eq.~\ref{Rep}) for the evaluation of the dispersion relation, based on the lack of transport feedback on the roughness length found in the velocity profile (Fig.~S5). However, we do use Eq.~(\ref{Rep}) for the calculation of the particle Reynolds number in the comparison with field data (shown below), where, due to the larger grain size,  hydrodynamically smooth conditions are no longer satified.

Reassuringly,  the model also reproduces the wavelength and growth rates of dunes measured in the field\cite{lu2021direct} (Table~S4), as shown in Fig.~4, where we corrected both the dune growth rate and wavelength for the ambient-air wind-tunnel conditions. This was done by multiplying the dune wavelength measured in the field $\lambda^m_{D,\rm field}$ by the ratio of the predicted dune wavelength for the ambient-air wind-tunnel conditions $\lambda^p_{D,\rm wind-tunnel}$ to the field conditions $\lambda^p_{D,\rm field}$:
\begin{equation}
  \lambda^m_{D,\rm wind-tunnel} = \lambda^m_{D,\rm field} \left(\lambda^p_{D,\rm wind-tunnel}/\lambda^p_{D,\rm field}\right),
\end{equation}
where the dune wavelength is defined as the fastest growing wavelength without subscale saturation ($\gamma = 0$). Similarly, the dune growth rate measured in the field $\sigma^m_{D,\rm field}$ was first rescaled using Eq.~(\ref{sigma+}) for field conditions (Table~S4) to get $\sigma^{+m}_{D,\rm field}$, and then multiplied by the ratio of the dimensional dune growth rate predicted for the ambient-air wind-tunnel conditions $\sigma^p_{D,\rm wind-tunnel}$ and the dimensionless dune growth rate predicted for field conditions $\sigma^{+p}_{D,\rm field}$:
\begin{equation}
  \sigma^m_{D,\rm wind-tunnel} = \sigma^p_{D,\rm wind-tunnel} \left(\sigma^{+m}_{D,\rm field}/\sigma^{+p}_{D,\rm field}\right).
\end{equation}
Note that in both expressions, the quality of the comparison relies on the model reproducing well the data for field conditions. All measured and estimated quantities used in the validation of the model are summarized in Table~S5. 


\begin{addendum}
 \item [Data availability] All data generated supporting the findings of this study are either extracted from other studies\cite{Duranetal19, miller1987wind,lu2021direct} or generated from our own measurements and can be made available upon request from the authors. Some of the data is also summarized within the Supplementary Information file (Tab. S1-S5).

 \item [Code availability] The code that integrates the equations of the hydrodynamic bedform model and the grain-scale numerical simulation used for this study can be made available upon request from the authors.


\end{addendum}


\end{document}